\documentclass[11pt,a4paper]{article}

\usepackage{graphicx}% Include figure files
\usepackage{epsfig}
\usepackage[dvips]{color}

\usepackage{hyperref}
\usepackage{graphicx,epsf}
\usepackage{bm}
\usepackage{amsmath}

\newcommand{\vecb}[1]{{\bf #1}}
\newcommand{\fpar}[2]{\frac{\partial{#1}}{\partial{#2}}}

\begin{document}

\begin{center}
{\large{\textbf{Numerical validation of the electromagnetic gyrokinetic code NEMORB\\
on global axisymmetric modes.}}}\\
\vspace{0.2 cm}
{\normalsize {A. Biancalani, A. Bottino, Ph. Lauber and D. Zarzoso\\}}
\vspace{0.2 cm}
\small{Max-Planck Institute for Plasmaphysics, 85748 Garching, Germany \\
www.ipp.mpg.de/$\sim$biancala}
\end{center}

\begin{abstract}
This is a report about a comparison of collisionless simulations on global modes (i.e. low poloidal mode number) with gyrokinetic code NEMORB against analytical theory and other codes. Only axisymmetric modes, i.e. with toroidal mode number n=0, are considered, and flat equilibrium profiles. Benchmarks are performed for GAMs against local analytical theory. In the presence of energetic ions, local benchmarks of NEMORB are performed against semilagrangian gyrokinetic code GYSELA. The models of adiabatic vs trapped-kinetic- vs fully-kinetic-electrons and of electrostatic vs electromagnetic at very low beta are compared. Scalings of Alfv\'en modes are also presented.
\begin{center}submitted to {\it Nuclear Fusion}\end{center}
\end{abstract}

\section{Introduction}

Understanding the dynamics of plasma instabilities in tokamaks is important because they are deleterious for the stability itself and because their interaction with energetic particles (EP) can redistribute the population of EP and reduce the efficiency of external plasma heating. 
Here we use the gyrokinetic code NEMORB to study collisionless simulations of  global (i.e. small ploloidal mode number) axisymmetric (toroidal mode number n=0) modes in tokamaks, both in absence and in the presence of EP. We consider flat equilibrium profiles, and compare with local analytical theory and with gyrokinetic code GYSELA. The code NEMORB~\cite{Bottino11} is a global gyrokinetic PIC code, derived as the multispieces, electromagnetic version of the code ORB5~\cite{Jolliet07}, which was written for studies of turbulence in tokamak plasmas.  The code GYSELA is a global flux-driven electrostatic semilagrangian gyrokinetic code~\cite{Grandgirard08,Sarazin10}. Axisymmetric modes like geodesic acoustic modes (GAMs)~\cite{Winsor68} have been found to play a role in the regulation of 
turbulence in ASDEX Upgrade~\cite{Conway11}. In addition, energetic GAMs 
(EGAMs)~\cite{Fu08,Zarzoso12} have been observed to significantly modified and modulate the 
turbulence in gyrokinetic simulations~\cite{Zarzoso13}.

This first section of this paper is an introduction and description of motivations for our work.

In the second section we provide the model equations of NEMORB.

In the third section, we focus on simulations of the Geodesic Acoustic Mode (GAM), that is a mode with m=0 n=0 potential perturbation and m=1 n=0 density perturbations, present in tokamak plasmas due to geodesic curvature, and with mainly electrostatic polarization. A GAM is observable in numerical simulations by initialising a zonal perturbation in ion density with a radial gradient. The resulting radial electric field creates an ExB drift along the poloidal direction, which starts oscillating due to the nonhomogeneity given by tokamak curvature. This oscillation, named GAM, is damped by Landau damping and it decays leaving a residual zonal radial electric field that is called residual zonal flow. The frequency and damping of the linear GAM oscillation is calculated at each radial position analytically~\cite{Winsor68,Zonca96,Zonca98,Sugama06,Gao06,Sugama07,Hallatschek07,Zarzoso12} and compared with numerical results of NEMORB (see also Ref.~\cite{
Angelino08,Vernay10}). We perform this 
comparison for flat density and temperature profiles, and nearly flat q profiles, as a benchmark of NEMORB against local analytical theory. For the same simulations, residual flow comparisons with analytical theory are also made.
%{\color{blue}{
Different models to treat the electrons are compared, and electromagnetic simulations at very low beta are performed, as a benchmark for electromagnetic version of NEMORB.
%}}

In the fourth section, we add an EP (in particular energetic deuterium) population. In this case the frequency and damping rate of GAMs is modified and a new branch named Energetic-ion GAM (EGAM)~\cite{Fu08,Zarzoso12} appears together with GAMs, with different frequency and growth/damping rate. Due to the similarity in spatial structure, the problem of how classifying GAMs and EGAMs arises: we define for this particular EP shifted-Maxwellian distribution function, EGAMs as those whose growth rate increases with energetic particle concentration. Local benchmarks of NEMORB on EGAMs against GYSELA are performed. Different models to treat the electrons are compared, and electromagnetic simulations at very low beta are performed, as a benchmark for electromagnetic version of NEMORB. 

In the fifth section, we focus on modes with mainly Alfv\'en polarization. We show first comparisons of numerical vs analytical scalings for Alfv\'en waves with $n=0$.

Finally, in the sixth section we give conclusions and describe our future work.

\section{Model equations of NEMORB}

NEMORB is a global nonlinear gyrokinetic $\delta f$ particle-in-cell PIC code~\cite{Bottino11}, derived as the multi-specie electromagnetic versionof ORB5~\cite{Jolliet07}. The Lagrangian formulation that is used, is based on the gyrokinetic (GK) Vlasov-Maxwell equations of Hahm, Brizard and Sugama~\cite{Hahm88,Brizard07,Sugama00}. The Lagrangian and Hamiltonian are given by:
\begin{equation}
L = \left(e
\vecb{A}+p_z\vecb{b}\right)\cdot\dot{\vecb{R}}+\frac{m^2}{e}\mu \dot{\theta} - H
\end{equation}
\begin{equation*}
H = m\frac{U^2}{2}+ m\mu B+ eJ_0\Phi+\mathcal{O}(\Phi^2)
\end{equation*}
where $\mu ={v_\perp^2}/{2B}$, $mU = p_z-e J_0 A_\parallel$ and $U=\fpar{H}{p_z}$.
Here $(\vecb{R},p_z,\mu,\theta)$ are the particle coordinates, $U$ is the parallel velocity of the particle and $J_0$ is the gyroaverage operator. The equations of motion of NEMORB are the Euler-Lagrange equations:
\begin{equation}
 \dot{\vecb R}=\fpar{H}{p_z}\frac{\vecb{B^*}}{B^*_\parallel}-\frac{1}{eBB^*_\parallel}\vecb{F}\cdot\nabla H~~~~~~\dot{p_z}=-\frac{\vecb{B^*}}{B^*_\parallel}\cdot\nabla H
\end{equation}
%\aeq
%\vecb{F}=\nabla\vecb{A}-(\nabla \vecb{A})^T~~~~\vecb{F}=\epsilon\cdot\vecb{B}~~~~~\nabla \times \vecb{b} = -\nabla\cdot\frac{\vecb{F}}{B}~~~~\vecb{b} \times = -\frac{\vecb{F}}{B}\cdot
%\eeq
Replacing the Hamiltonian in the Euler-Lagrange eqs. we obtain:
\begin{align*}
\dot{\vecb R}&=\left(\frac{p_z}{m}-\frac{e}{m}J_0A_\parallel\right)\frac{\vecb{B^*}}{B^*_\parallel}+\frac{1}{B^*_\parallel}\vecb{b}\times \left[\mu\frac{m}{e} \nabla B + \nabla J_0\Psi\right]\\
\dot{p_z}&=-\frac{m\vecb{B^*}}{B^*_\parallel}\cdot\left[\mu \nabla B + \frac{e}{m} \nabla J_0\Psi\right]\\
\Psi &\equiv \Phi - \frac{p_z}{m} A_\parallel
\end{align*}
where $\vecb{B^*}=\nabla\times\vecb{A^*}$, $\vecb{A^*} \equiv \vecb{A}+(p_z/e)\vecb{b}$ and $\vecb{F}=\epsilon\cdot\vecb{B}$.
Gyrokinetic Vlasov equations are solved for ions, drift-kinetic equations for electrons. The Vlasov equation is: 
\begin{equation}
\frac{{\rm d} f}{{\rm d}t}=\fpar{f}{\vecb{R}}\cdot \dot{\vecb R} + \fpar{f}{p_z}\dot{p_z}=C(f)+S
\end{equation}
%$f(\vecb{R},p_z,\mu)$, distribution function of the gyrocenters; $C(f)$, collision operator, $S$, source term.\\
where $C(f)$ is the collision operator, and $S$ is the source term.
Electron-ion collisions are retained, and self-collisions are not used here. In the case of the simulations described in this paper, we have collisionless dynamics ($C(f)=0$) without sources ($S=0$), and energy and momentum conservation can be proved via gyrokinetic field theory~\cite{Scott10}.

Finally the linearized gyrokinetic Poisson equation (in the long wavelength approximation) and parallel Amp\'ere's law are:
%\begin{align*}
\begin{eqnarray}
-\nabla_\perp\left(\sum_{\rm species}\frac{mn_0}{eB^2}\right)\nabla_\perp\Phi&=&\sum_{\rm species}~\delta n\\
\left(\sum_{\rm species} \frac{\mu_0n_0e^2}{m} A_\parallel \right)-\nabla_\perp^2A_\parallel&=&\mu_0 \sum_{\rm species} \delta j_\parallel
\end{eqnarray}
%\end{align*}
%$\delta n=\int {\rm d}W\delta f \delta(\vecb{R}+\boldmath{\rho} -\vecb{x})$,
%$\delta j_\parallel=e\int {\rm d}W(p_z/m)\delta f \delta(\vecb{R}+\boldmath{\rho} -\vecb{x})$, ${\rm d}W=2\pi B^*_\parallel m^{-1} {\rm d}\vecb{R} {\rm d}p_z {\rm d}\mu$.\\

The boundary conditions on the fields are unicity condition  (solution does not depend on  magnetic angle $\chi$, where $\chi = (1/q) \int^\theta B\cdot\varphi  /  B\cdot\theta' d\theta'$) and 0 radial derivative at radius $s=0$ and $\phi=A_\|=0$ at radius $s=1$. Regarding particles, they are reflected with $\chi=-\chi$ when they exit the flux surface with $s=1$. 

Ideal MHD equilibria (via the CHEASE code~\cite{Luetjens96} can be used as an input.
Noise reduction and dissipation is performed via a modified Krook operator or coarse graining (not used here). NEMORB is a multi-ion species code, which means that impurities or fast particles can be initialized as initial state and let evolve like a separate specie. NEMORB is also massively parallelized.

\section{Geodesic acoustic modes (GAMs)}

% 
% \begin{figure}[b!]
% \begin{center}
% \includegraphics[width=0.45\textwidth]{loc-GAM-q_s}
% \caption{\label{fig:loc-GAM-equil} Example of q-profile for the GAM case with q = 2 (s is the radial coordinate, $s\simeq r/a$). Nearly flat q-profiles (i.e. with low shear) are chosen for these simulations in order to investigate the local limit.}
% \end{center}
% \end{figure}

\subsection{Equilibrium and simulation parameters}

We choose an analytical tokamak equilibrium with circular flux surfaces and high aspect ratio ($\varepsilon=a/R=0.1$), toroidal major radius $R_0 =1.3 m$, and toroidal magnetic field $B_{tor} = 1.9 T$. We consider flat temperature and density profiles. Very low values of shear are considered ($\hat s \sim 10^{-2}$), so that the q profiles are nearly flat, and our simulations can be compared with local analytical theory. We initialize a deuterium charge density perturbation of the form $\sin(\pi r/a)$, that has a radial gradient but is independent of the poloidal and toroidal angle, and we let it evolve linearly.

Typical simulations have a spatial grid of  (s,$\chi$,$\phi$) = 64x64x4 and a time step of 10-100 $\Omega_i^{-1}$, with $10^6 - 10^7$ markers. The length of a typical simulation is $2\cdot 10^5 \, \Omega_i^{-1}$, corresponding to 2000-20000 time steps. GAMs oscillations are observed, and we measure the radial electric field amplitude, frequency and damping rate.

\begin{figure}[t!]
\begin{center}
\includegraphics[width=0.42\textwidth]{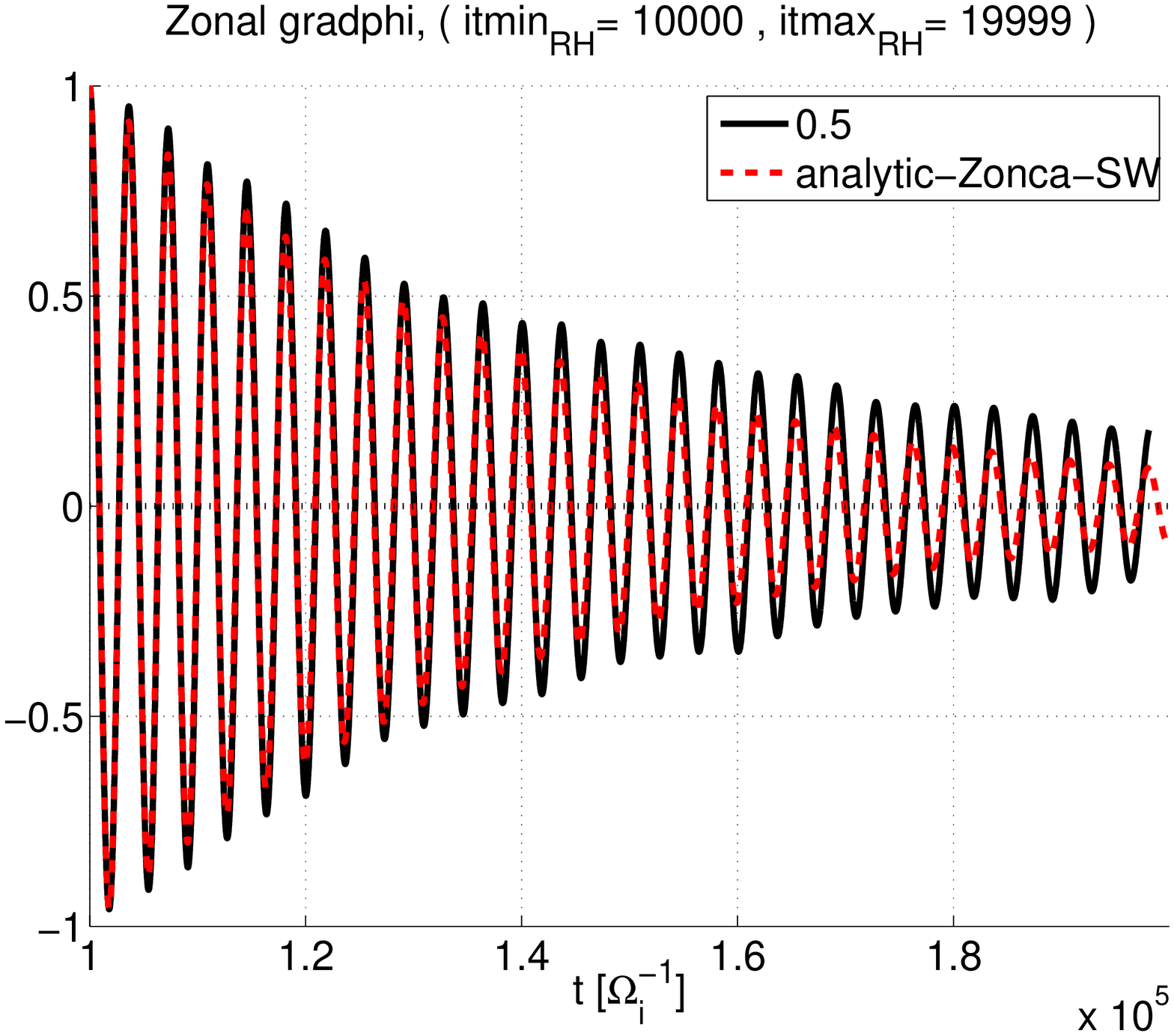}
\includegraphics[width=0.42\textwidth]{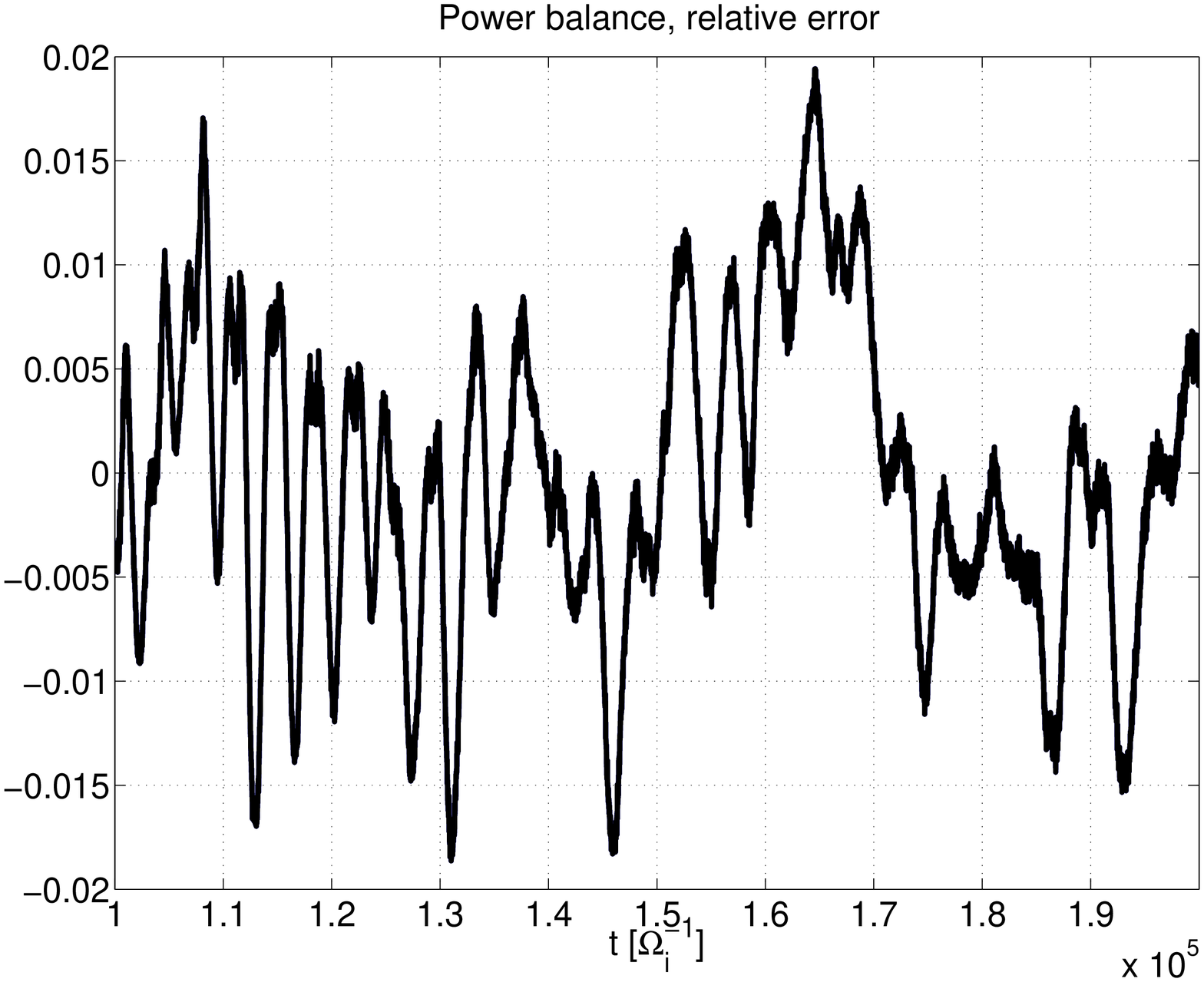}
\caption{\label{fig:loc-GAM-RH_run-check} On the left: radial electric field oscillation vs time, measured with NEMORB near the center of a simulation box, at radius $s \simeq 0.5$, for $\rho^* = \rho_s/a= 1/160$. The residual has been subtracted here. In dashed red line, the analytic prediction with frequency given by Zonca96~\cite{Zonca96} and damping rate given by the explicit formula of Sugama07~\cite{Sugama07} is also shown. On the right, the relative error of the power balance for the same simulation. Here a zoom in t= ($10^5$, $2 \cdot 10^5 )\, \Omega_i^{-1}$ is shown. This simulation is centered at q=1.5 and has $\tau_e=1$. Its  spatial grid is  (s,$\chi$,$\phi$)=64x64x4 points and the time step is 10 $\Omega_i^{-1}$, with $10^7$ markers.}
\end{center}
\end{figure}

An example of the evolution in time of the amplitude of the radial electric field (i.e. the radial gradient of the flux-surface-averaged  scalar potential) is shown in Fig.~\ref{fig:loc-GAM-RH_run-check}, measured at a location with radius $s\simeq 0.5$. In the same figure, we plot the function $\cos(\omega_{GAM} * t)\exp (-\gamma_{GAM} * t)$, where we take $\omega_{GAM}$ from analytic theory of Ref~\cite{Zonca96} and $\gamma_{GAM}$ from Ref.~\cite{Sugama07}. We find a good fit of the analytical and numerical radial electric field evolution in time, shown there in a time range after the relaxation of the initial perturbation, which takes a few GAM oscillations.

\subsection{Power balance check}

In order to quantify the numerical errors affecting the results of our linear simulations, we use here a diagnostic that measures the power balance between the perturbed field growth of the mode and the power transferred  from the particles (see Ref.~\cite{Bottino04}). The former is measured by calculating the time derivative of the energy contained in the perturbed field for a certain wave number k, $dE_{field}/dt$. The latter, is calculated as $dE_k/dt = j\cdot E_k$. The relative error of the power balance, $\delta E$, is calculated and gives us how much the power balance is violated:
\begin{equation}
\delta E_{rel} = \frac{ dE_{field}/dt - dE_k/dt}{dE_{field,max}/dt}
\end{equation}
where $dE_{field,max}/dt$ is the maximum value of $dE_{field}/dt$ in the time interval of interest. In Fig.~\ref{fig:loc-GAM-RH_run-check}, the relative error of a NEMORB simulation is shown. For this specific run, that falls within 2\%.

\begin{figure}[t!]
\begin{center}
\includegraphics[width=0.44\textwidth]{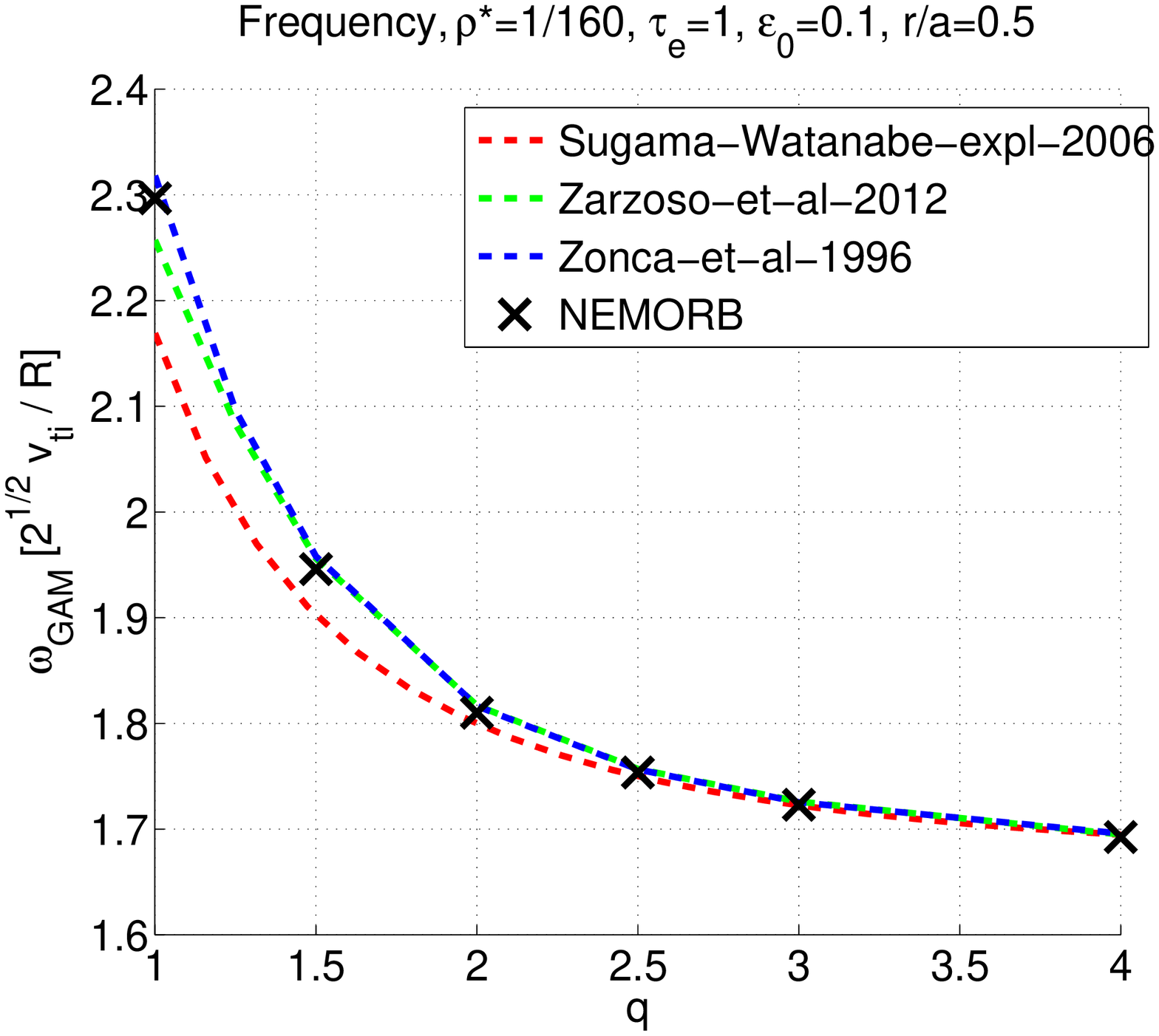}
\includegraphics[width=0.42\textwidth]{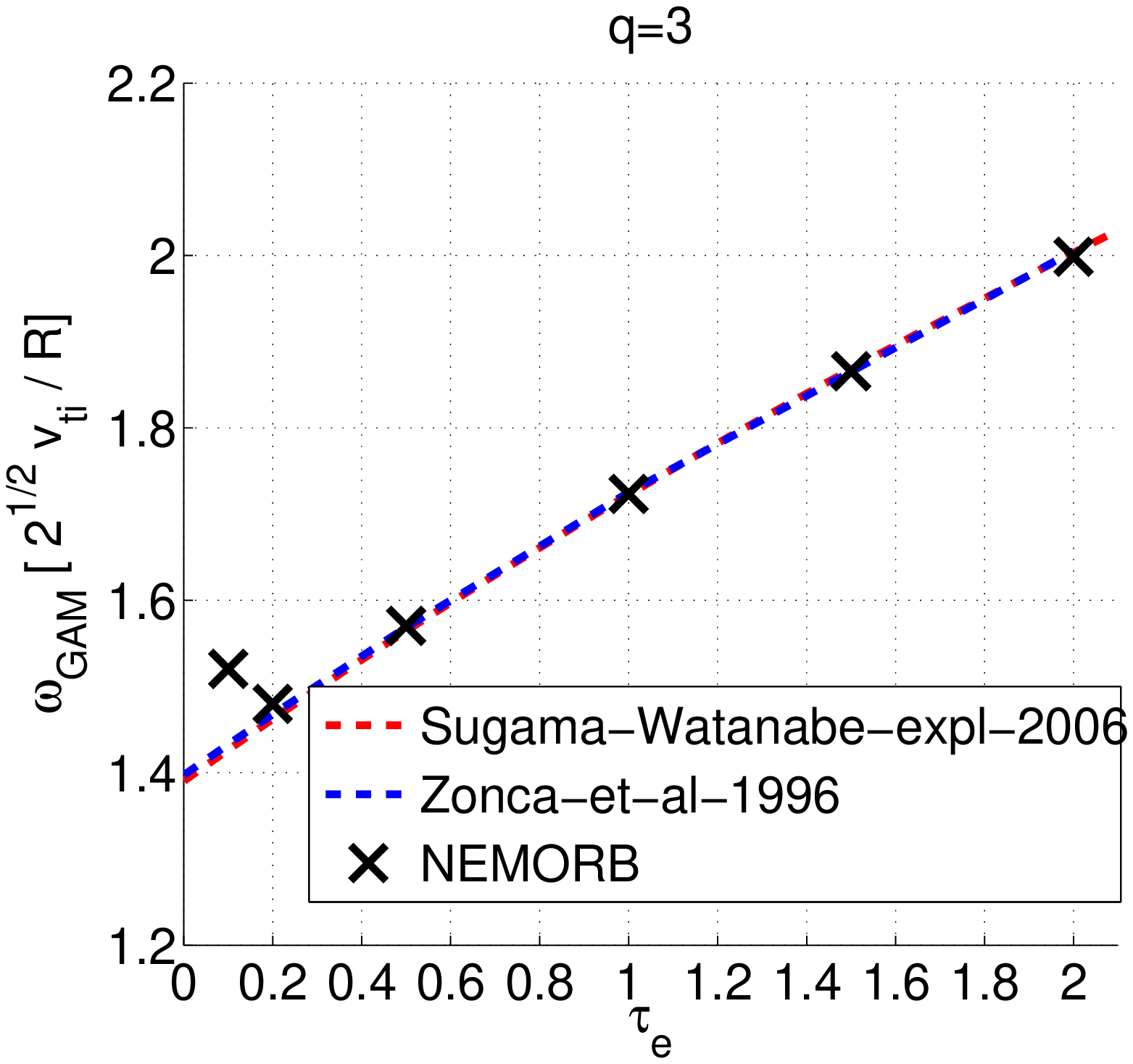}
\caption{\label{fig:loc-GAM-omega-q_omega-taue} On the left: GAM frequency for several local simulations with $\tau_e = 1$, centered at different values of q. Values given by NEMORB (black) fit well with the analytic theory of Ref.~\cite{Zonca96} (blue) and with approximated theory of Ref.~\cite{Zarzoso12} (green) at all values of q. Explicit formula (Eq.~2.9 of Ref.~\cite{Sugama06}) valid for large q is also shown in blue. On the right: GAM frequency for several local simulations with $q=3$ and different values of $\tau_e$. A good agreement is found except at very low values of $\tau_e$.}
\end{center}
\end{figure}

\subsection{Frequency}

Here we show two benchmarks: the first with the scalings of the GAM frequency vs the safety factor q and the second of the GAM frequency vs the electron to ion temperature ratio $\tau_e$. A plasma with $\rho^* = \rho_s/a = 1/160$ is considered.
For the first benchmark, several simulations are carried out, all with $\tau_e = 1$, and each with different q profile. Each q profile is linear and centered at a rational surface with a particular value of q. 
The frequency measured with NEMORB for the several simulations with different q-profiles is found to agree well with gyrokinetic dispersion relation given in Ref.~\cite{Zonca96} (see Fig.~\ref{fig:loc-GAM-omega-q_omega-taue}):
\begin{equation}\label{eq:GFLDR}
i \Lambda = \delta W_f + \delta W_k
\end{equation}
where $\Lambda$ is the generalized inertia (defined in Appendix~\ref{sec:appendix}), $\delta W_f$ is the ideal MHD contribution to the potential energy perturbation and $\delta W_k$ is the kinetic component to $\delta W$ due to the energetic particle compressions. In our cases we have $\delta W_f = \delta W_k = 0$.
An approximated electrostatic relation given by Ref.~\cite{Zarzoso12} is also shown to fit well with NEMORB results. The explicit formula given by Eq.~2.9 of Ref.~\cite{Sugama06} is derived for large values of q, and consistently it is shown to fit well in the appropriate limit.  For the second benchmark, several simulations are performed, all with q=3, and each with different value of $\tau_e$. The comparison with analytic theory is shown in Fig.~\ref{fig:loc-GAM-omega-q_omega-taue}. We obtain a good agreement for this scaling except at very low values of $\tau_e$, where further study is needed.

\subsection{Damping rate}

\begin{figure}[t!]
\begin{center}
\includegraphics[width=0.46\textwidth]{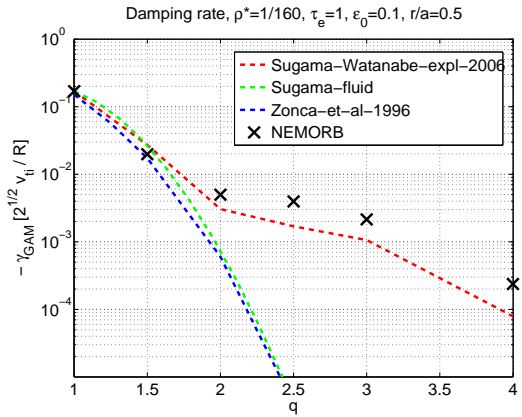}
\caption{\label{fig:loc-GAM-gamma_q} Damping rate for the same simulations as in Fig.~\ref{fig:loc-GAM-gamma_q}. Each simulation has the q profile centered at a different value. Results of NEMORB (black crosses) are compared with explicit analytical formula of Sugama07 (red dashed line), with the same formula where the FOW corrections have been neglected (green dashed line) and with the dispersion relation of Zonca96 (blue dashed line). We can see that for these simulations the FOW effects are more and more important the more we go to higher values of q.}
\end{center} 
\end{figure}

The damping rate of GAMs can also be measured in NEMORB and compared with analytical theory. We find that this study is more numerically demanding than the study of frequencies: for low number of markers the amplitude of electric field oscillation is observed not to decrease exponentially, due to numerical errors. Therefore we increase the number of markers up to $10^8$ and decrease the time step down to $dt = 10 \, \Omega_i^{-1}$. We find that the scaling of NEMORB results is in good agreement with analytic theories for $q<2$ (see Fig.~\ref{fig:loc-GAM-gamma_q}), but for larger values of q finite orbit width (FOW) effects become important, and the theories not accounting for them are observed to deviate. These FOW effects are found to be difficult to estimate. In fact, the initial electric field perturbation develop during the evolution large radial gradients especially near the border $s=1$ (see also Fig.~\ref{fig:loc-GAM-residuals}), due probably to our boundary conditions, and this 
increases the value of 
$k_\perp \rho_i$ and affects the dynamics 
also at lower radii. We measure $k_\perp$ and use it in the explicit formula of Sugama07~\cite{Sugama07}, obtaining a qualitative good match in the scaling with q. Further comparison at $q>2$ is on the way. As a next step on this field, further work is needed to increase the accuracy or decrease $k_\perp$ of our perturbation. Other theories (like Zonca98~\cite{Zonca98}) are also going to be compared with NEMORB's results.

\subsection{Zonal flow residuals}

The residual zonal electric field, that remains after the GAM oscillation is damped in collisionless simulations, can be measured and compared with analytical prediction. Rosenbluth and Hinton in 1998~\cite{Rosenbluth98} calculated the residual zonal flow $A_r = u_{E} (t=\infty) / u_E (t=0)$ (with $u_E$ being the ExB drift), for small values of inverse aspect ratio $\epsilon = r/R_0$ and in the limit of radial wave-length much larger than the poloidal gyroradius $\rho_p = \rho_i q/\epsilon$:
\begin{equation}
A_{r,RH} =  \frac{1}{ 1 + 1.6 \, q^2 / \epsilon^{1/2} }
\end{equation}
Lateron, in 2006, Xiao and Catto extended this formula for radial wave-length of the same order of magnitude as the poloidal gyroradius, and added higher-order corrections of aspect-ratio~\cite{Xiao06}:
\begin{equation}\label{eq:residuals_XC}
A_{r,XC} = \frac{1}{ 1 +  q^2 \, \Theta / \epsilon^2 }
\end{equation}
with $\Theta$ defined as:
\begin{displaymath}
 \Theta = \Big( 1.6 \, \epsilon^{3/2} +\frac{1}{2} \, \epsilon^2 + 0.36 \, \epsilon^{5/2} \Big) - 2.44 \, \epsilon^{5/2} \, k_\perp^2\rho_p^2
\end{displaymath}

\begin{figure}[t!]
\begin{center}
\includegraphics[width=0.47\textwidth]{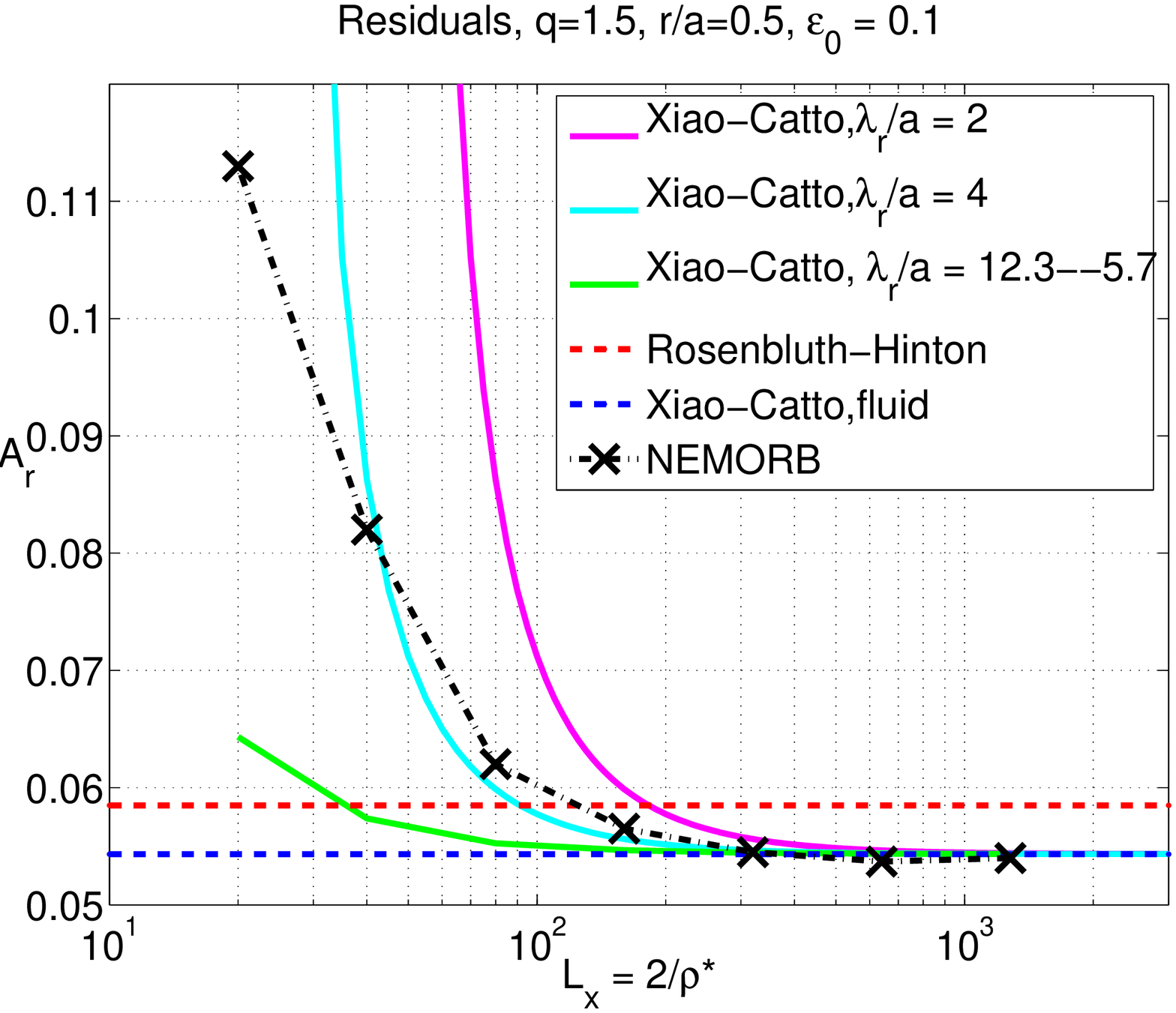}
\includegraphics[width=0.40\textwidth]{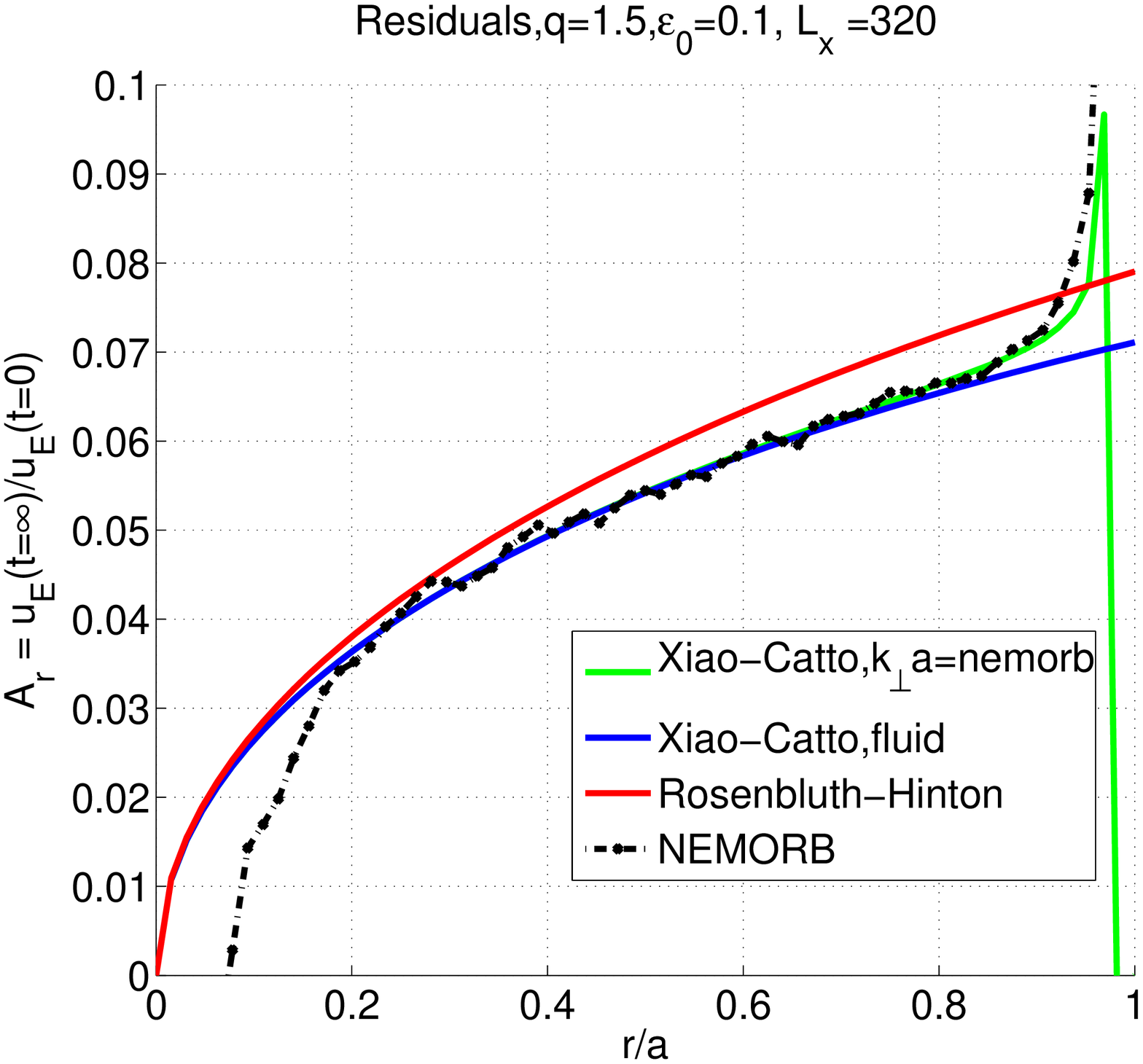}
\includegraphics[width=0.42\textwidth]{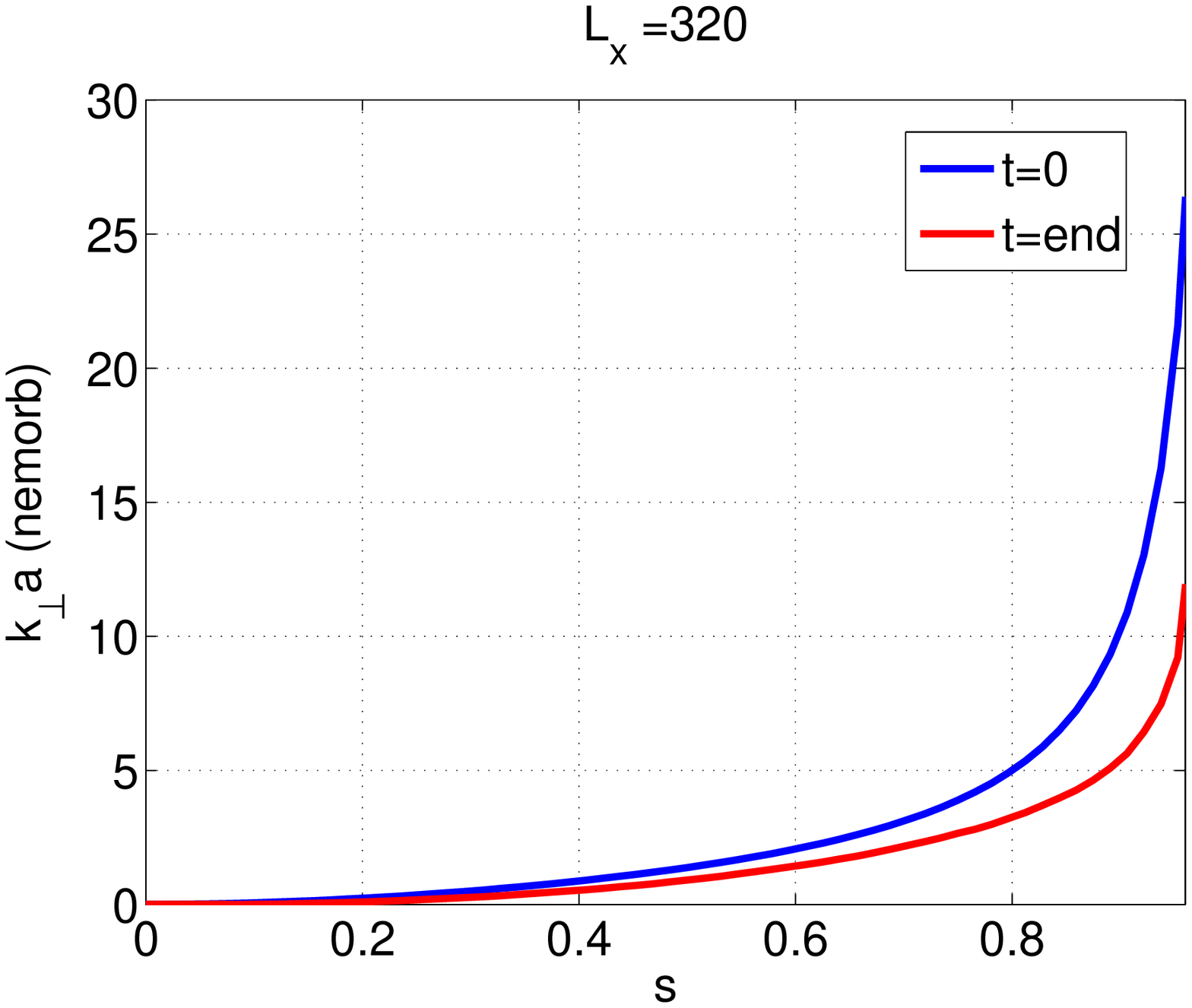}
\caption{\label{fig:loc-GAM-residuals} Residuals of the zonal flow $A_r$  for an equilibrium with $q=1.5$, $\epsilon_0 =0.1$.  On the top left, the fluid limit is studied by repeating the same simulation with increasing values of the parameter $L_x = 2/\rho^* = 2 a/\rho_s$. In this limit, the results of simulations with NEMORB (black points) tend to the fluid limit of the analytical prediction of Xiao and Catto (blue dotted horizontal line). The Rosenbluth-Hinton value is also given as a red dotted horizontal line. Prediction of Xiao and Catto for finite poloidal gyroradius $\rho_p = \rho_i q/\epsilon$ are also shown, for values of $\lambda_r = 2a$, $\lambda_r = 4a$ and $\lambda_r$ measured by NEMORB at the end of the simulation (which returns $\lambda_r /a \in [12.3, 5.7]$). On the top right, the profile of the residuals is shown for the simulation with $L_x = 320$. The difference of the residuals at $s=r/a=0$  from the Xiao-Catto prediction is thought to be due to the boundary conditions. In the lower 
graph, the 
value of $k_\perp a$ 
measured 
by 
NEMORB is shown, at the beginning and at the end of the simulation with $L_x = 320$.}
\end{center}
\end{figure}

In Fig.~\ref{fig:loc-GAM-residuals} the residual flow is shown for simulations with $q=1.5$ and $\epsilon_0 =0.1$. We measure it in NEMORB as the average of $\nabla \bar\phi (t) / \nabla\bar \phi(t=0)$ over the last few periods of GAM oscillation (here $\bar\phi$ is the flux-surface-averaged scalar potential). The fluid limit is studied by repeating the simulations with smaller and smaller value of $\rho^*$. The values of $L_x$ are chosen in the set $[20 \, ; \, 1280]$, correspondoing respectively to $\rho^* \in [0.1 \, ; \, 0.0016]$. This limit is also called local limit in gyrokinetic simulations of turbulence~\cite{Candy04}.
For values of $L_x$ larger than 200, corresponding to values of $\rho^*$ smaller than 0.01, the results of NEMORB are in good agreement with the fluid limit of the analytical prediction of Xiao and Catto, namely Eq.~\ref{eq:residuals_XC} where we neglect the finite poloidal Larmor-radius effects (expressed in the term with $k_\perp \rho_p$). On the other hand, for values of $\rho^*$ larger than 200, we still do not find a good quantitative agreement of NEMORB's results with the analytical prediction of Xiao and Catto with finite-Larmor-radius effects included. The analytical predictions of Xiao and Catto shown in Fig.~\ref{fig:loc-GAM-residuals} have been calculated by using three different values of $\lambda_r / a$. The first two values are $\lambda_r / a = 2$ and $\lambda_r / a = 4$, where $\lambda_r / a = 2$ is the value which is consistent with the radial shape of our initial perturbation in density ($\delta n \sim \sin (\pi r/a)$), but it does not fit NEMORB'
s results, so we have 
plotted also $\lambda_r / a = 4$ for comparison. These two values correspond respectively to $k_\perp \rho_p \in [9.42 \, ; \, 0.147] $ and $k_\perp \rho_p \in [4.71 \, ; \, 0.074] $. The third value of $\lambda_r/a$ used for the calculation of the prediction of Xiao and Catto is in fact a profile in s, derived from the value of $k_\perp a$ measured in NEMORB as $\nabla \bar\phi / \bar \phi$ over the last few periods of GAM oscillation. This profile is shown for a simulation with $L_x = 320$ in Fig.~\ref{fig:loc-GAM-residuals} and compared with the profile measured at $t=0$. This simulation has a grid of (64,64,4) points in the (s,$\chi$,$\phi$) direction, a time step of dt=10 $\Omega_i^{-1}$ and $10^7$ markers. The discrepancy we have between the theory and numerical results for simulations with $L_x < 320$ might be due to the fact that we do not measure correctly the value $k_\perp \rho_p$ in our simulations, or to the fact that the analytical prediction 
is derived in a local theory whereas our 
simulations are global. As next steps in this direction, this discrepancy needs to be understood better, for example the link of $k_\perp a$ measured at the beginning and the initial perturbation density.

\begin{figure}[b!]
\begin{center}
\includegraphics[width=0.47\textwidth]{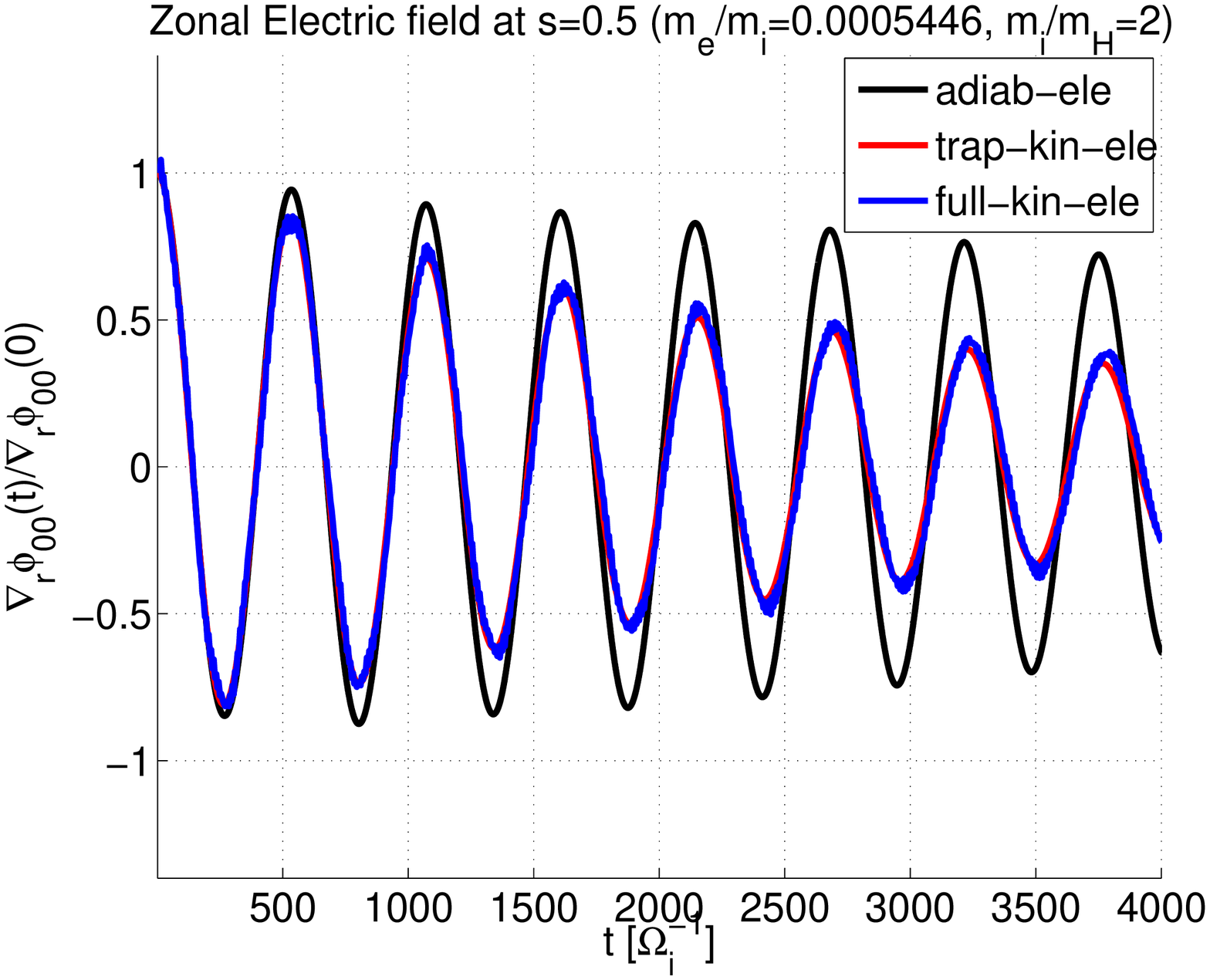}
\includegraphics[width=0.48\textwidth]{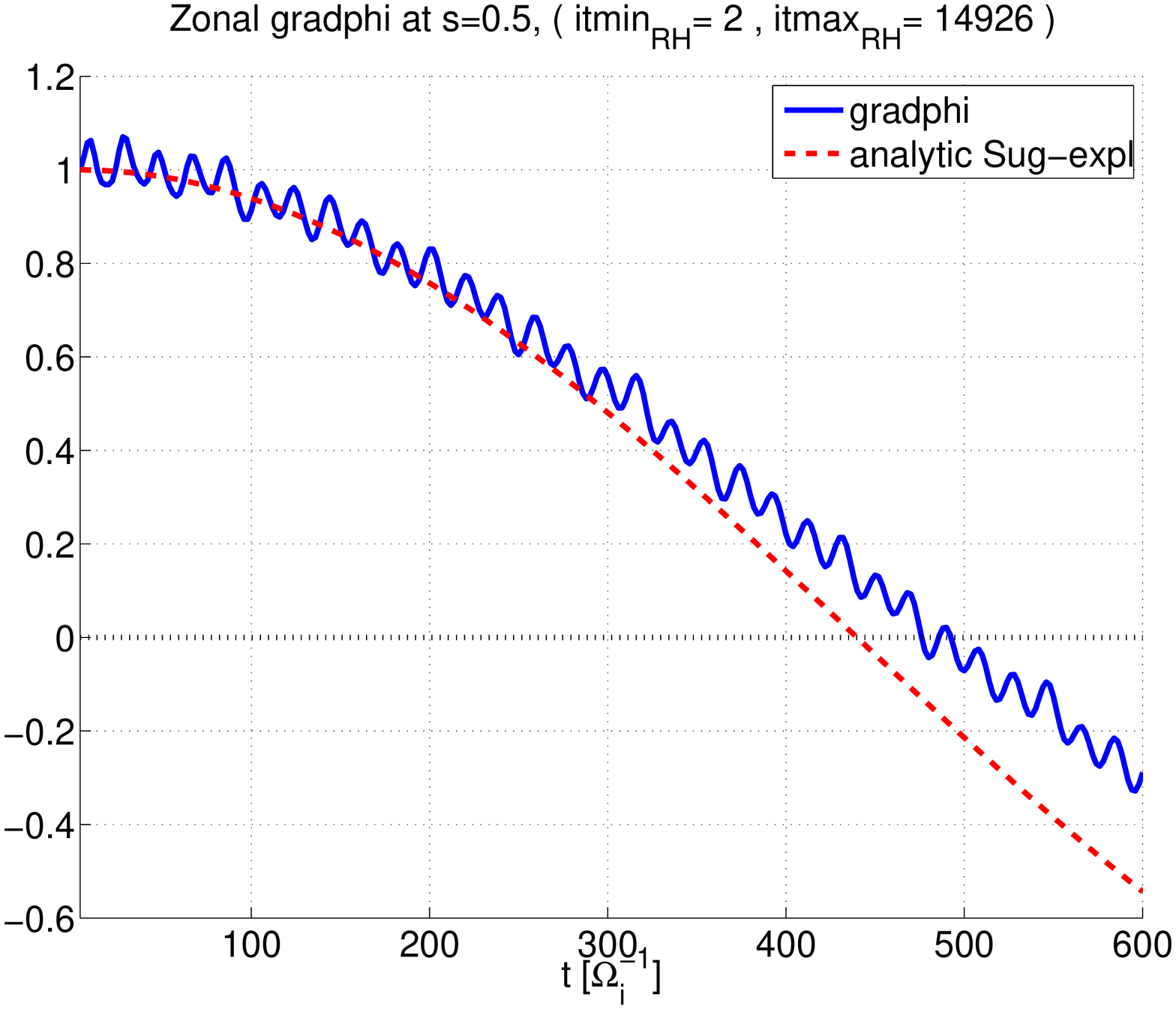}
\caption{\label{fig:GAM-trap-full-ele} On the left: radial electric field oscillations of GAMs for three different simulations with adiabatic electrons (black), trapped-kinetic-electrons (red) and fully-kinetic-electrons (blue). The adiabatic electrons model differs from the other two ones, only for a smaller damping rate. On the right, zoom in time for the fully-kinetic-electrons model, compared with analytic prediction of GAM oscillation. Higher frequency oscillation are a fictitious effect of the model, due to the dynamics of passing electrons when density tends to zero.}
\end{center}
\end{figure}

\subsection{From adiabatic electrons to kinetic electrons}

In the previous section, Rosenbluth-Hinton simulations with adiabatic electrons have been described. Now we want to describe the differences we find, if we treat the electron species with a kinetic model, rather than adiabatic. We have two models for this treatment: the trapped-kinetic-electrons model and the fully-kinetic-electrons model.
In the trapped-kinetic-electrons model, NEMORB selects at t=0 the electrons which will perform trapped orbits and treat them kinetically, whereas those which will perform passing orbits are treated adiabatically. The advantage of this model is that wave-particle resonances are treated for trapped electrons, and modes like trapped electron modes are included in the model, and at the same time NEMORB does not need to resolve the parallel electron dynamics, which requires a very small time step. 
In the fully-kinetic-electrons model, all electrons are treated kinetically, like ions. This model has the advantage that also wave-particle interaction with passing electrons are retained in the model, but the time step has to be set small enough to resolve the fast parallel electron dynamics.

In Fig.~\ref{fig:GAM-trap-full-ele}, we can see the evolution of the radial electric field oscillation of a GAM, for the three different models of electrons of NEMORB. As we can see, frequency does not change much in the three models, whereas damping rates are higher when trapped electrons are treated kinetically, possibly due to resonances with bounce frequencies of trapped electrons (consistently with Ref.~\cite{Zhang10}).

\begin{figure}[b!]
\begin{center}
\includegraphics[width=0.47\textwidth]{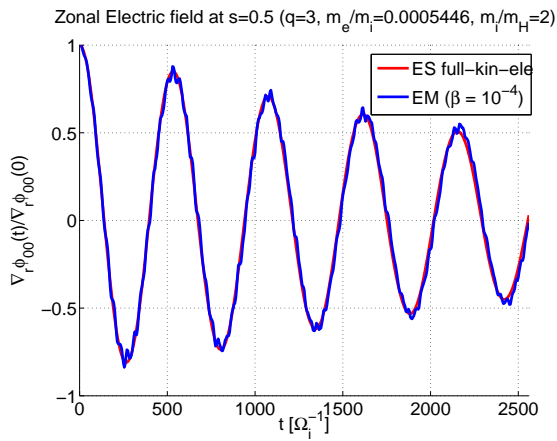}
\caption{\label{fig:GAM-elemag} Rosenbluth-Hinton benchmark of NEMORB in electromagnetic mode with $\beta=10^{-4}$ against NEMORB in electrostatic mode. We can observe that the dynamics of the GAM oscillation is matching well, except for the presence of higher frequency shear Alfv\'en wave dynamics at higher frequency, in the electromagnetic mode.}
\end{center}
\end{figure}

When we make a zoom in time of the GAM oscillation, when the electrons are treated kinetically, we observe a higher frequency oscillation (see Fig.~\ref{fig:GAM-trap-full-ele}). This high frequency oscillation  is a spurious  result of the model, in the sense that it is not present in nature but it is consistent with the particular model we are using. It is the limit of kinetic Alfv\'en waves for $\beta$ going to zero (electrostatic model) at fixed temperature~\cite{Lee83,ScottPC}:
\begin{displaymath}
V_A^2 = v_A^2 (1 + k_\perp^2 \rho_s^2)/(1 + k_\perp^2 d_e^2) 
\end{displaymath}
\begin{center}
$\downarrow \; (\beta_e \rightarrow 0$ at const. T)\\
$\downarrow \;\;\;\;\;\;\;\;\; (v_A^2 = v_e^2 \mu_e /\beta_e $)
\end{center}
\begin{displaymath}
V_A^2 = v_e^2 (1 + k_\perp^2 \rho_s^2)/(k_\perp^2 \rho_s^2) 
\end{displaymath}
\begin{displaymath}
\hookrightarrow \omega \sim (v_e/2\pi \rho_s) (a/R)
\end{displaymath}
This is a pure result of the model: consistent treatment of finite beta eliminates this purely artificial wave.

\subsection{From electrostatic to electromagnetic at low beta}

We now show results of Rosenbluth-Hinton simulations in the case where the electrons are treated kinetically and NEMORB also solves the Ampere's law. In this case, also magnetic perturbations are treated in the model. At this stage of code validation, we restrict our simulations to very low values of beta $\beta=10^{-4}$, and we compare with simulations where Ampere's law is not solved (electrostatic mode). The equilibrium is a local equilibrium with $q=3$
%(\emph{comment: add more details about the equilibrium?})
.

As we can see in Fig.~\ref{fig:GAM-elemag}, the benchmark of electromagnetic vs electrostatic mode of NEMORB is succesfull for these Rosenbluth-Hinton tests: in fact the dynamics is found to be the same. If we make a zoom in time, we find that higher frequency oscillations are present together with the GAM oscillation. These are shear-Alfv\'en oscillations (with toroidal mode number n=0), that are initialized implicitly in the Rosenbluth-Hinton test due to finite $\epsilon$ effects (which couples different m-modes). These Alfv\'en modes will be described more in details in Sec.~\ref{sec:SAW}.

\section{Energetic-ion driven GAMs}
\label{sec:EGAM}

\begin{figure}[b!]
\begin{center}
\includegraphics[width=0.45\textwidth]{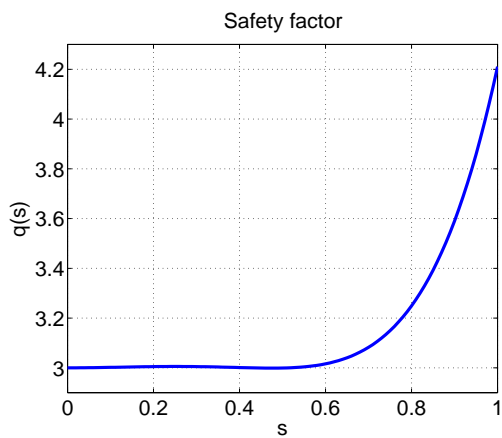}
\caption{\label{fig:safety-factor} Safety factor used in our simulations. For most of the radial domain the value of the safety factor is $q\simeq 3$. In GYSELA we restrict only to the radial domain $0.2<s<0.8$.}
\end{center}
\end{figure}

In this section we describe results of GAMs with energetic particles, for equilibrium profiles of the simulations performed with GYSELA~\cite{Grandgirard08,Sarazin10} and described in Ref.~\cite{Zarzoso12}: these are flat equilibrium temperature and density profiles and a nearly flat q profile. They are expected to behave as local simulations, due to the fact that the profiles are nearly flat. Results of nonlinear electrostatic simulations are shown, with electrons treated adiabatically. In previous section, NEMORB has been succesfully benchmarked on GAMs against analytical theory for linear simulations falling into this limit (see also Ref.~\cite{Angelino08,Vernay10}). In particular, in this report we show some first results of simulations with NEMORB where also an energetic particle population is present. At this step we consider with NEMORB a bump-on-tail fast particle population.

The dependence, in the linear phase, of frequency and growth/damping rates of GAMs and EGAMs on fast particle concentration is shown, and compared with results of GYSELA shown in Ref.~\cite{Zarzoso12}.

\subsection{Equilibrium and simulation parameters}

We choose a tokamak equilibrium with circular flux surfaces and moderate  aspect ratio ($R = 1 m$, $\varepsilon=a/R=0.3125$), $B=1.9 \, T$, 
with flat equilibrium temperature and density profiles (we choose $\tau_i = T_i/T_e = 1$ and $\rho^* = \rho_s/a = 1/64$). Very low values of shear are considered at the center of the simulation box, where the measurements are done, so that the q profile is nearly constant with value $q=3$, and our simulations can be considered local (see Fig.~\ref{fig:safety-factor}). In the inner half of the radial domain the value of the safety factor is always very near $q=3$, with a maximum around $s=0.25$ (s is the squared root of the poloidal magnetic flux, used here as radial coordinate). In the outer half of the radial domain, the safety factor increases to reach the value of $q=4.2$ at the edge.

%\\ \emph{comment: to be compared with simulations with very flat safety factor profile.}\\
We initialize at $t=0$ a charge density perturbation of the form $\sin(\pi r/a)$, that has a radial gradient but is independent of the poloidal and toroidal angle. In NEMORB, a bump-on-tail distribution function for the fast particle population is implemented at the initialization of our simulations, $t=0$, like in Eq.~27 of Ref.~\cite{Zarzoso12}:

\begin{figure}[b!]
\begin{center}
\includegraphics[width=0.50\textwidth]{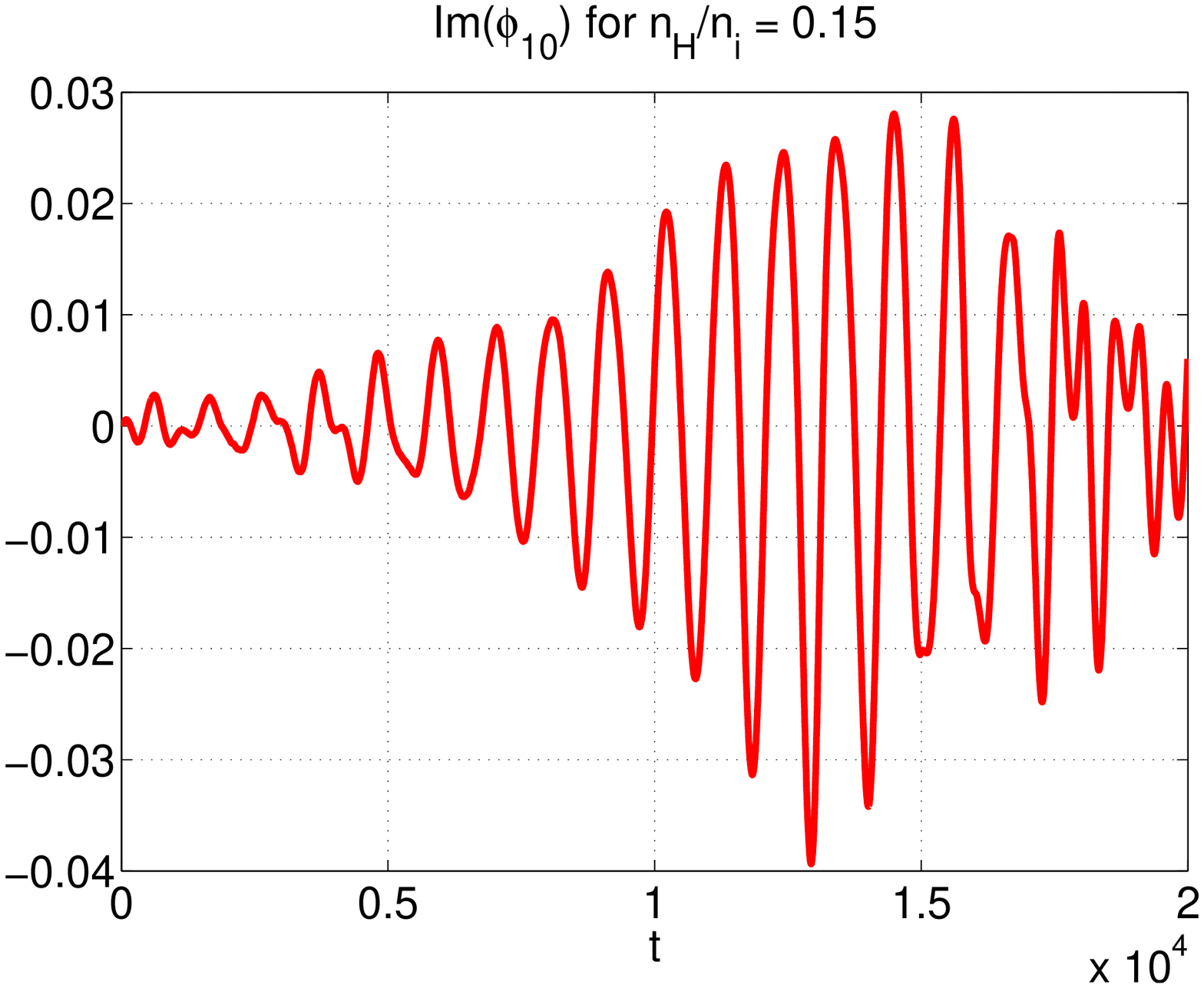}
\includegraphics[width=0.475\textwidth]{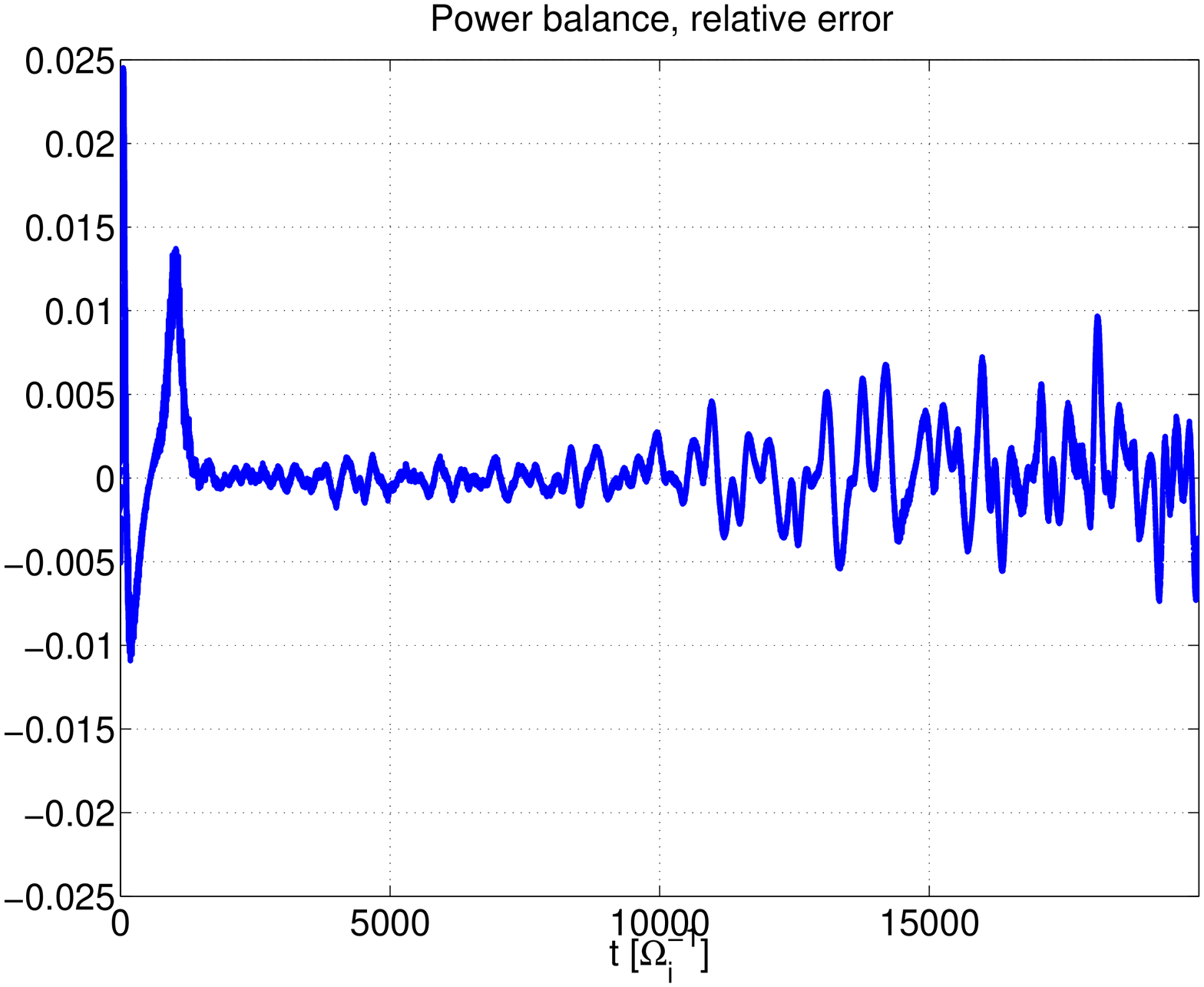}
\caption{\label{fig:RH_run-check} On the left: imaginary part of the complex Fourier transform in $\theta$ of the zonal component (n=0) of the scalar potential $\phi$, measured at the center of the radial domain, $s=0.5$, vs time
%(\emph{comment: t in units of $\Omega_i^{-1}$})
. On the right, the relative error of the power balance for the same simulation.}
\end{center}
\end{figure}

\begin{equation}
F_{eq,h} = F_{M,h} e^{-\frac{\bar\zeta^2}{2 \hat{T}_h}} \cosh \Big( \frac{\bar\zeta \, \zeta}{\hat{T}_h} \Big) 
\end{equation}
where $F_{M,h} = \frac{n_h}{2\pi T_i \hat{T}_h/m} e^{-\frac{\zeta^2 + 2\bar\mu}{2 \hat{T}_h}} $, and  $\zeta = \sqrt{ 2(E - \mu B_{min})/m } /v_{th}$. We choose the hot ion normalized temperature as $\hat{T}_h = T_h / T_i = 1$ and the normalized mean parallel velocity of hot ions as  $\bar\zeta = v_\| /v_{ti} = 4$.
The perturbation is let evolve in time in a nonlinear electrostatic simulation with adiabatic electrons. We are interested here only in the linear phases.

%\\ \emph{comment: to be compared with linear simulations}\\
A typical simulation has a spatial grid made of (s,$\chi$,$\phi$)=64x64x4 points and the time step is 0.5~$\Omega_i^{-1}$, with $10^7$ markers. The length of a typical simulation is $2*10^4~\Omega_i^{-1}$, corresponding to $4*10^4$ time steps. Running in parallel on 512 cpu it takes about 15 hours.

\subsection{Power balance check}

Before analyzing the result of the simulation, we check the conservation of energy and calculate an estimation of the numerical error, similarly to what is done also for GAMs in absence of EP in previous section.
In order to quantify the numerical errors affecting the results of our simulations, we use here a diagnostic that measures the power balance between the perturbed field growth of the mode and the power transferred  from the particles (Ref.~\cite{Bottino04}). The only difference with the diagnostic used for GAMs, is that here we take into account also the energy transfer with the energetic particles. In Fig.~\ref{fig:RH_run-check}, the relative error of a NEMORB simulation is shown. Orientatively, for these simulations we consider as physical the dynamics in the time phase where this relative error falls within 1\%.

\begin{figure}[b!]
\begin{center}
\includegraphics[width=0.51\textwidth]{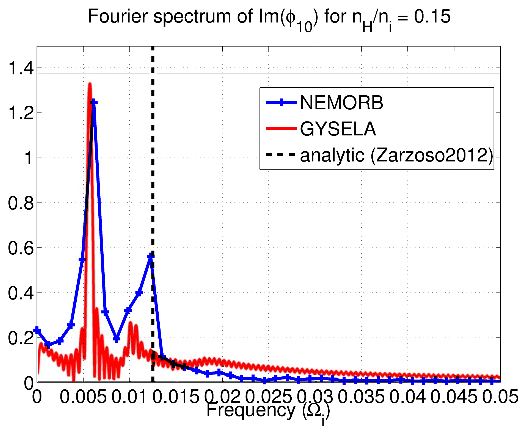}
\includegraphics[width=0.47\textwidth]{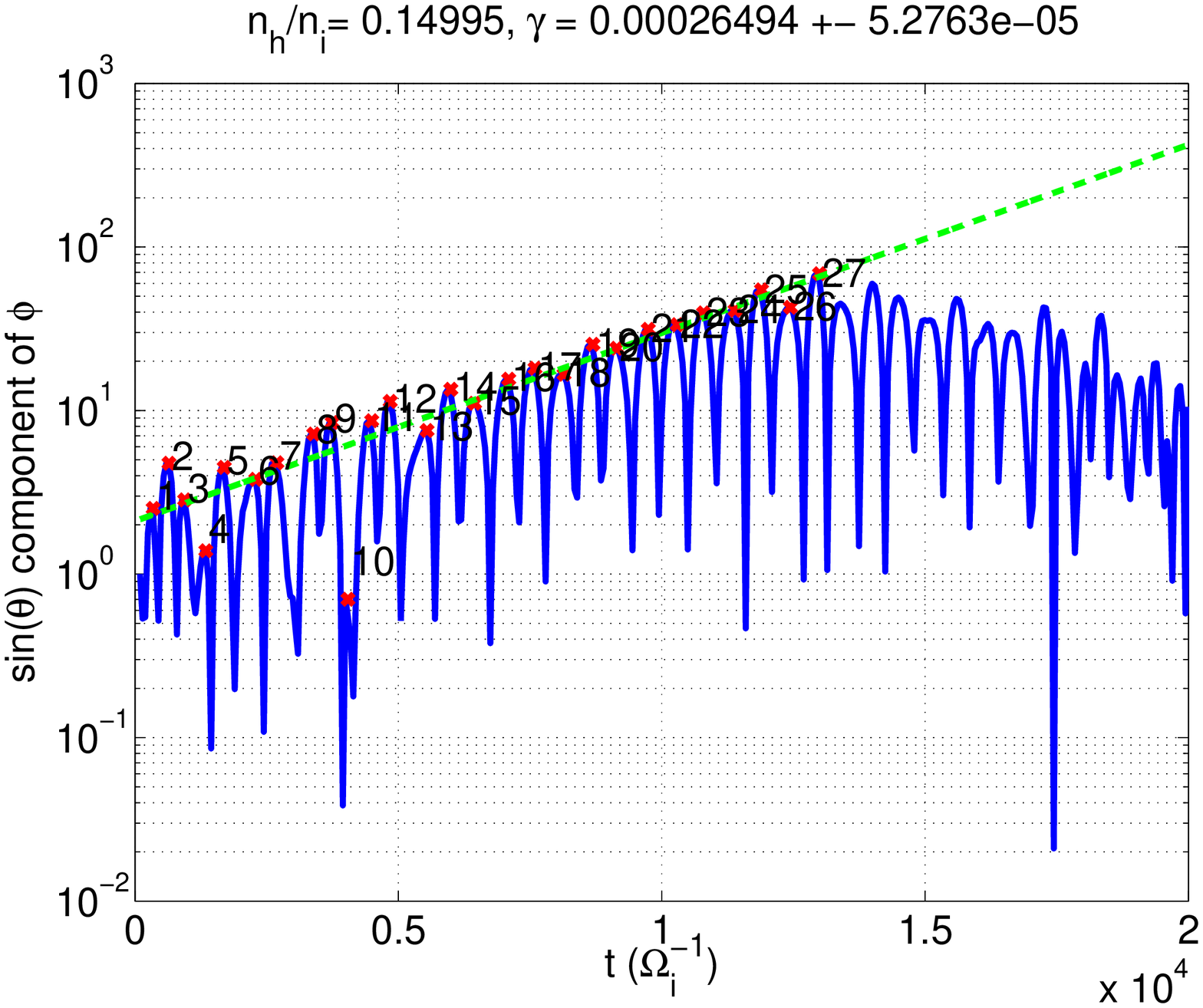}
\caption{\label{fig:Fourier} On the left, Fourier tranform in time of Im($\phi_{10}$) with NEMORB (blue points) compared with results obatined with GYSELA and described in Ref.~\cite{Zarzoso12}. On the right, absolute value of Im($\phi_{10}$) in logarithmic scale, vs time, with the linear fit performed to measure the growth rate.}
\end{center}
\end{figure}

\subsection{Frequencies and growth rates}

In a typical run we observe oscillations of Im($\phi_{10}$) (see Fig.~\ref{fig:RH_run-check}), that is the imaginary part of the complex Fourier transform in $\theta$ of the zonal component (n=0) of the scalar potential $\phi$ (in other words, Im($\phi_{10}$) is the Fourier coefficient of $\phi$ in $\theta$ relative to the $\sin(\theta)$ component). The typical period of the oscillations is of the order of the sound time $\tau_s \sim R/c_s$. We can observe the coexistence of two modes with different frequencies, where one is damped (GAMs) and one is growing (EGAMs). In the late phase we can observe a nonlinear saturation of the EGAM. In our simulations, we filter out all perturbations with toroidal mode number $n\ne 0$. Therefore, even though our simulations are nonlinear, no interaction with ITG or TEM turbulence is studied at this stage. 
The frequency of the modes is measured by performing a Fourier transform in time of Im($\phi_{10}$) at the mid radius, $s=0.5$. Two main frequencies are detected (see Fig.~\ref{fig:Fourier}): the higher frequency corresponds to GAMs and the lower to EGAMs. The growth rate of EGAMs is measured by performing a linear fit in logarithmic scale of the absolute value of Im($\phi_{10}$).

\begin{figure}[t!]
\begin{center}
\includegraphics[width=0.49\textwidth]{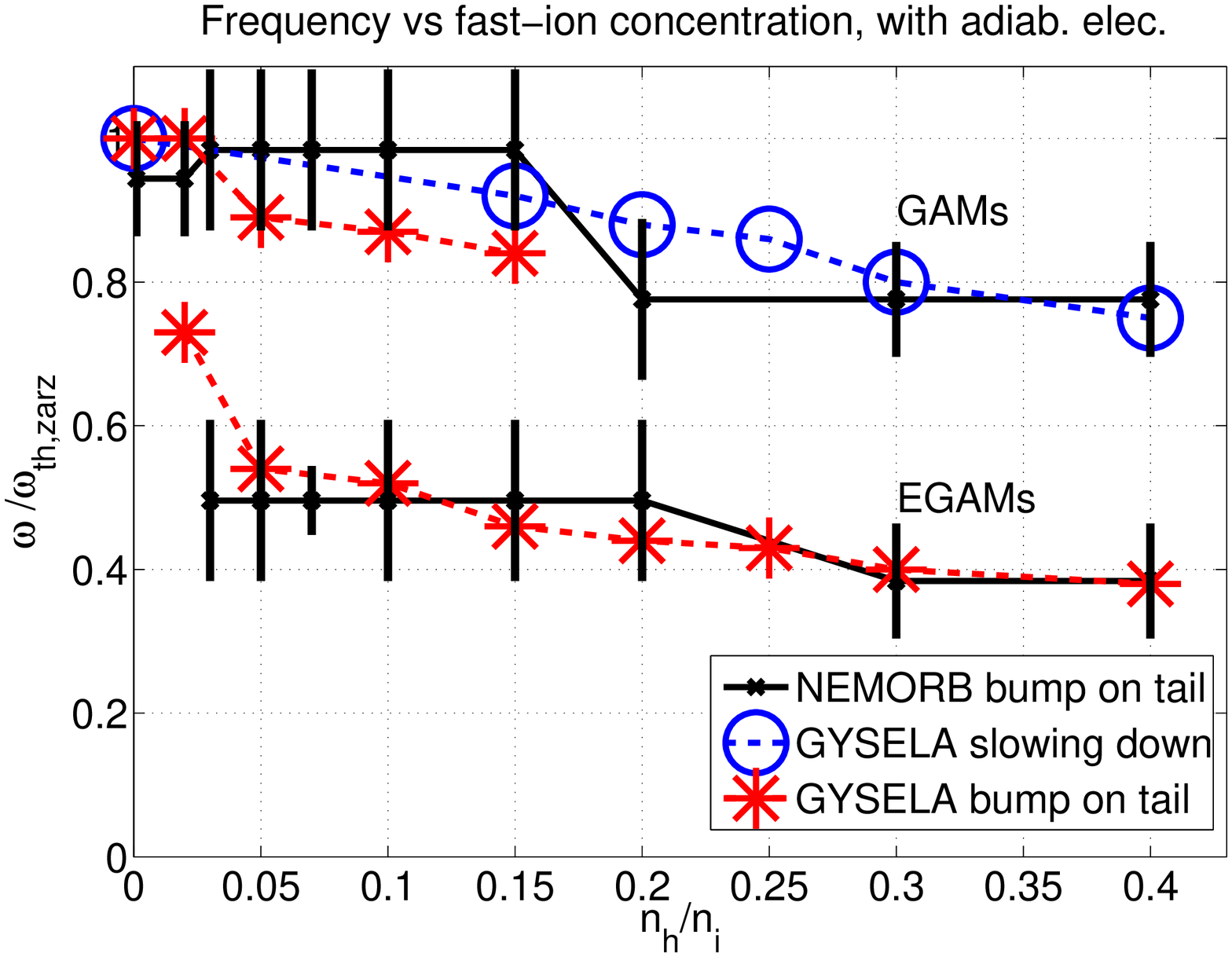}
\includegraphics[width=0.49\textwidth]{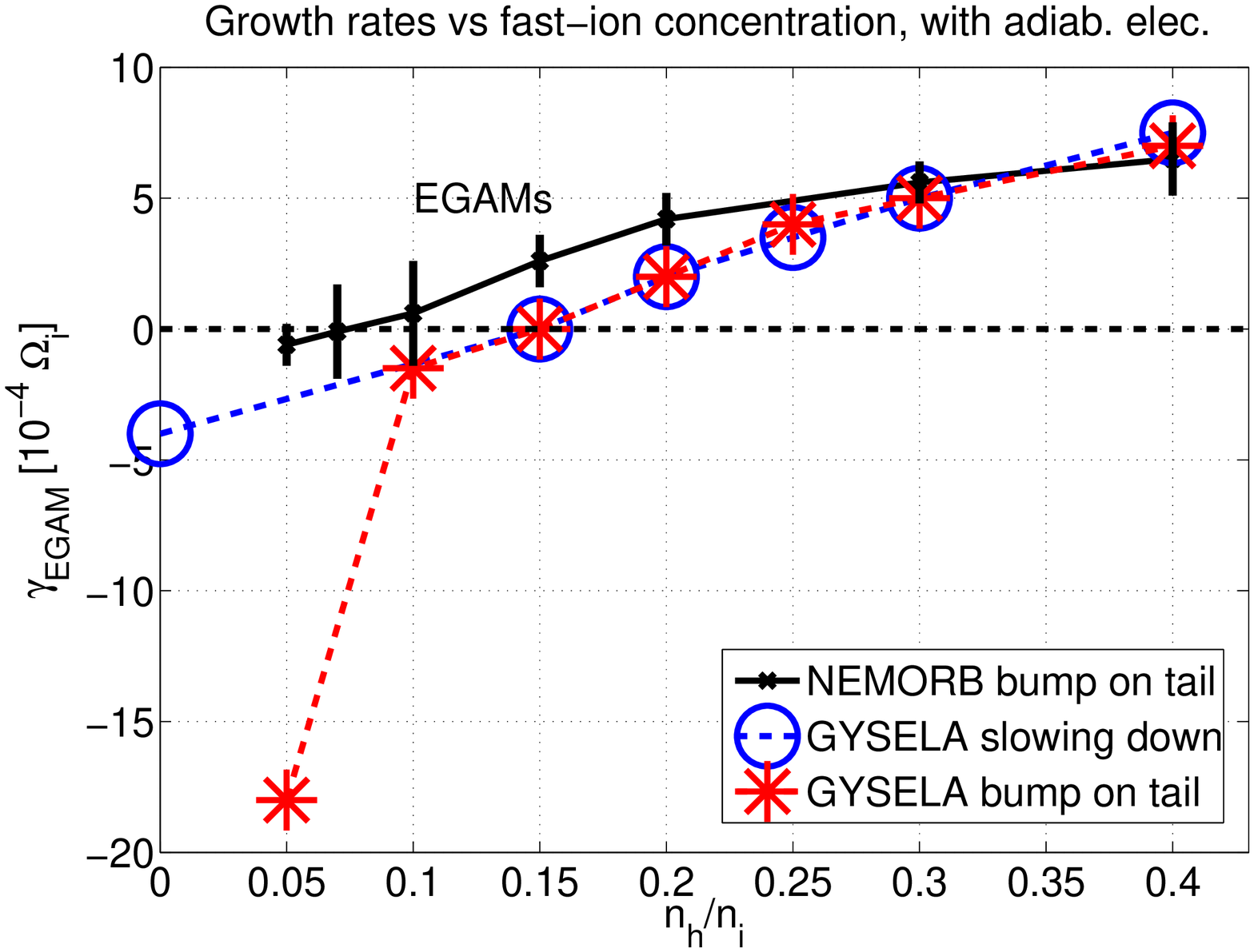}
\caption{\label{fig:omega-gamma} On the left: frequencies of GAMs (higher frequencies) and EGAMs (lower frequencies) vs fast ion relative concentration, normalized with the theoretical value given in absence of fast ions by Eq.~14 of Ref.~\cite{Zarzoso12}. The black points are measurement of simulations with NEMORB (only a bump-on-tail fast ion distribution function is presently implemented in NEMORB). Blue points and red points are results obtained with GYSELA and described in Ref.~\cite{Zarzoso12} respectively with slowing-down fast ion distribution function and bump-on-tail fast ion distribution function.
On the right: EGAM growth rate vs fast ion relative concentration. Measurement of damping rates GAMs is more difficult and is in currently in progress.}
\end{center}
\end{figure}

A scan of the frequency and growth rate of GAMs and EGAMs is shown in Fig.~\ref{fig:omega-gamma}, vs the relative EP concentration $n_h/n_i$. 
We can see that results of NEMORB agree qualitatively with those of GYSELA described in Ref.~\cite{Zarzoso12}. The large uncertainty of the frequency measurements of NEMORB is due to the fact that in our nonlinear simulations we have only a few oscillations in the scalar potential before the saturation. Since the saturation is a nonlinear interaction of wave and particles, we expect that by running linear simulations we won't have a physical saturation of the mode (but only a numerical one at later times) and therefore we expect to be able to run longer simulations and have more detailed measurements of the frequency.

%\\ \emph{comment for Alberto, is this true? comment by Philipp: do we have all non-linear terms switched off? otherwise we'll see a relaxation of the EP distribution function leading to saturation.
% We can also look at cases for weaker drive...}\\

Qualitative agreement of NEMORB with GYSELA is found also for the growth rates of EGAMs. Growth rates of EGAMs are observed to increase with fast ion concentration. The stability threshold measured with NEMORB is at the value of $n_h/n_i = 0.07$, to be compared with the threshold measured in GYSELA, of about $n_h/n_i = 0.15$.

\section{Shear Alfv\'en modes}
\label{sec:SAW}

\subsection{Equilibrium and simulation parameters}

\begin{figure}[b!]
\begin{center}
\includegraphics[width=0.47\textwidth]{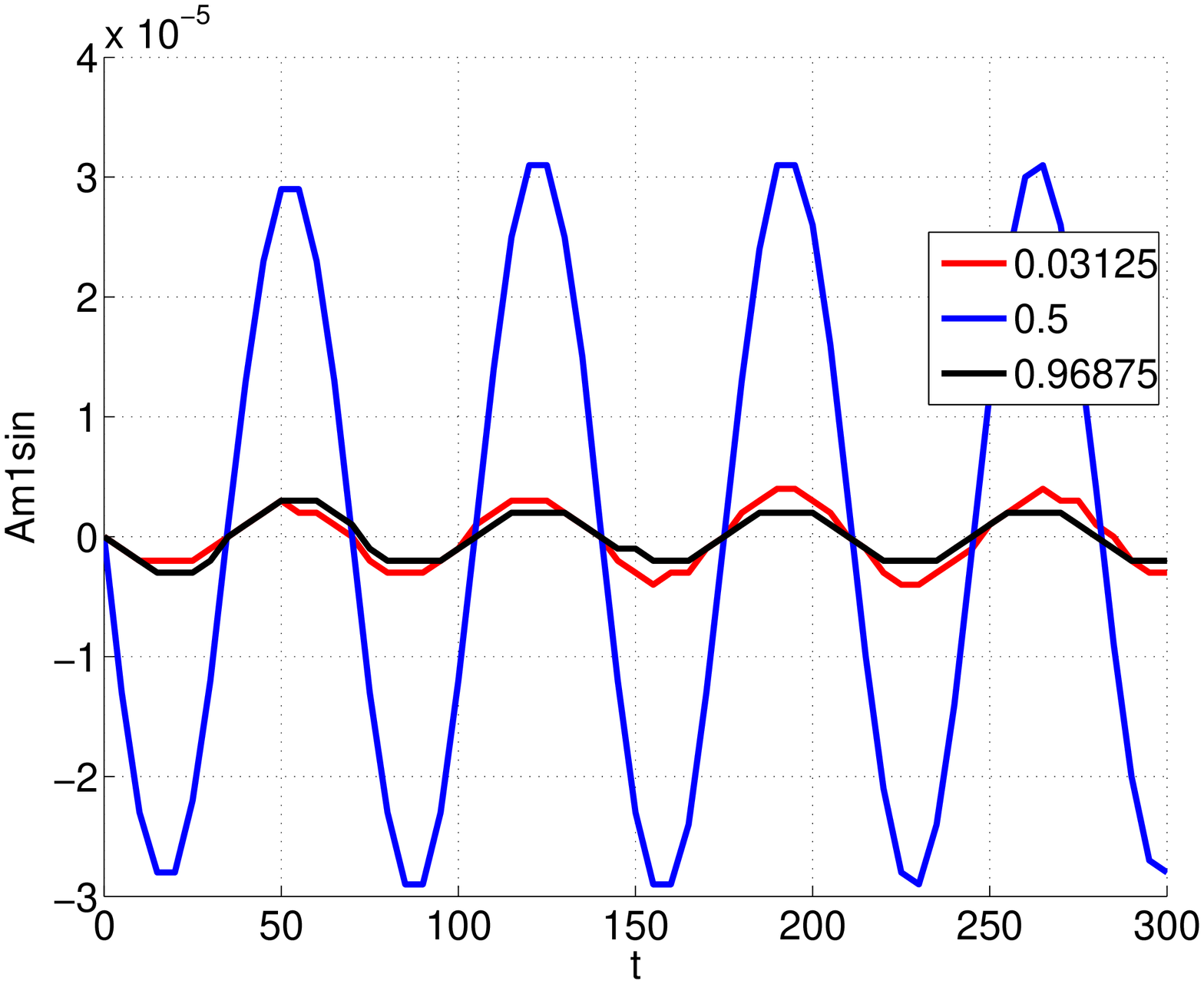}
\includegraphics[width=0.49\textwidth]{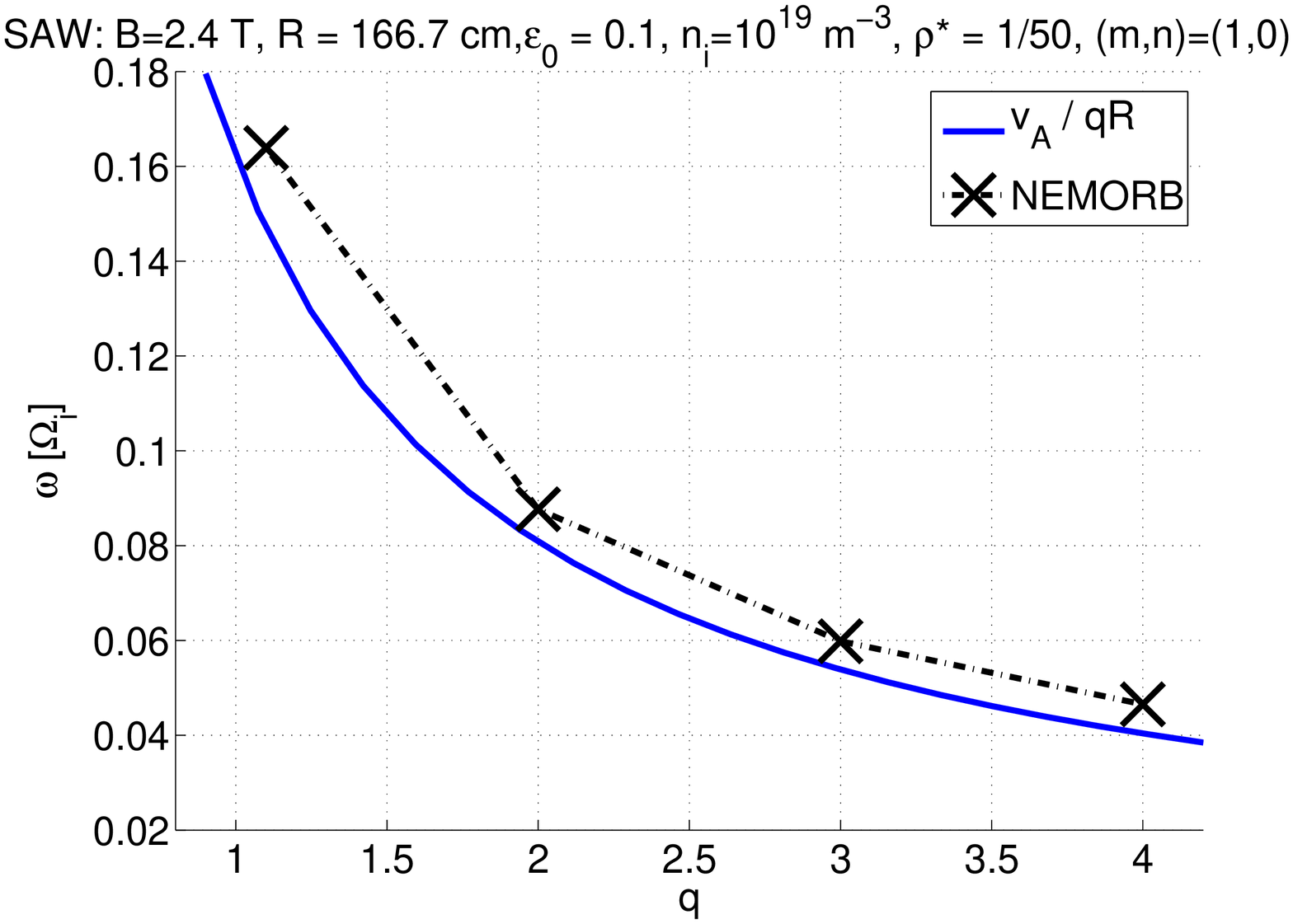}
\caption{\label{fig:SAW} On the left: Oscillation in time of the m=1 component of $A_\|$ ($\sin\theta$ part) at three different radial location. On the right, scan for different NEMORB simulations with different q profile (black points), compared with analytical scaling law (blue).}
\end{center}
\end{figure}

In order to perform these first scans of shear Alfv\'en modes with NEMORB, we choose a tokamak equilibrium with local profiles and large aspect ratio, in order to fall into the local regime. The major and minor radii are $R_0 = 1.667 m$ and $a = 0.1667 m$, with aspect ratio ($\varepsilon = 0.1$), circular cross section, toroidal magnetic field $B_{tor} = 2.4 T$. The density and temperature profiles are flat, with values $T_e = 1 keV$ and $\beta_e = 10^{-4}$, and we repeat different simulations with different q-profiles, each of them almost flat and centered at a different value of q. A typical simulation has a spatial grid of  (s,$\chi$,$\phi$) = 64x32x4 and a time step of 1 $\Omega_i^{-1}$, with $10^7$ markers for ions and $10^7$ for electrons. The length is $300 \, \Omega_i^{-1}$, corresponding to 300 time steps. The beginning of a GAM oscillation is observed in the perturbed scalar potential $\phi$, and on top of it we measure faster oscillations in 
parallel vector 
potential $A_\|$ (see Fig.~\ref{fig:SAW}).

%\subsection{Power balance check}
%to be written

\subsection{Scaling with q}

We measure the frequency of oscillation of the parallel component of the vector potential, $A_\|$, for different simulations with different q-profile.
Local theory of shear Alfv\'en wave predicts an oscillation with a frequency scaling like:
$ \omega = v_A k_\| = v_A m/qR_0 $, where $v_a = B/\sqrt{4\pi n m_i}$. In Fig.~\ref{fig:SAW}, the comparison is shown. We find a good qualitative agreement of NEMORB results with analytical scaling, which is the 1/q scaling (all other dependencies are kept constant in this scan). The reason why NEMORB's measured frequencies are slightly higher than analytical prediction might be the finite $\epsilon$ effect, and is to be investigated more in details.

\section{Conclusions and next steps}

In this paper, we have performed a numerical validation of the code NEMORB on GAMs in collisionless simulations with flat equilibrium profiles. Benchmarks against local analytical theory have been described for GAMs and against the code GYSELA for EGAMs. Scalings of frequencies, damping rates and electric field residuals have been analyzed. First scalings of Alfv\'en modes have also been shown.

In absence of energetic particles, we have found good matching of GAM frequency against analytical theory whereas the damping rates and the residuals have been found more difficult to compare, due to the strong dependence of finite orbit width effects on the radial wave vector $k_\perp$, which evolves in a global simulations leading to values different from the initial one. More detailed analysis is required in this context both on the numerical side, and on the analytical one, where we are going to compare also with other theories~\cite{Zonca98,Gao06}. Differences in the dynamics have been discussed as depending on the model used for treating electrons (adyabatic vs trapped-kinetic vs fully kinetic). Moreover, the first benchmark of GAMs between electromagnetic simulations at very low beta and electrostatic simulations has been showed.

In the presence of energetic particles, we have found a good agreement between frequencies obtained with NEMORB and with GYSELA. Regarding the growth rates, qualitative agreeement has been found between GYSELA and NEMORB, but quantitative differences are observed at low EP concentrations. Though this disagreement is still to be completely understood, among the possible explanations it we mention the difference in the boundary regions between both codes, since additional dissipative terms can be introduced in GYSELA in the boundary domain of the simulation box, whereas these terms are not considered in these NEMORB simulations.

Finally, first scalings of shear Alfv\'en modes have been showed as a validation of the code in electromagnetic modes for Alfv\'enic fluctuations.

As next steps, we are going to perform simulations of GAMs with realistic global tokamak equilibrium profiles, both in electrostatic and electromagnetic mode, and compare with other codes like LIGKA~\cite{Lauber07} (Linear Gyrokinetic Shear Alfv\'en). EGAMs are also going to be studied in electrostatic and electromagnetic mode, and the electromagnetic effects are going to be investigated. Alfv\'en modes in non-axisymmetric equilibria are going to be investigated, both in absence or in the presence of energetic particles.

%\begin{appendices}
 \appendix
\section{Generalized fishbone-like dispersion relation for GAMs}
\label{sec:appendix}

The generalized fishbone-like dispersion relation (GFLDR)~\cite{Zonca96}, as written in Eq.~\ref{eq:GFLDR}, describes linear electromagnetic oscillations in the presence of energetic particles, neglecting finite-Larmor-radii (FLR) effects and finite-orbit-width (FOW) effects, in the framework of gyrokinetic theory. For our purpose of benchmarking electrostatic simulations of GAMs in the absence of energetic particles, we can assume $\delta W_f = \delta W_k = 0$. The inertia term, $\Lambda$, is given by~\cite{Zonca96}:
\begin{equation}
\Lambda^2 (\omega) = \frac{\omega^2}{\omega^2_A} \Big(1-\frac{\omega_{*pi}}{\omega} \Big) + q^2 \frac{\omega \omega_{ti}}{\omega^2_A} \Big[ \Big(  1-\frac{\omega_{*ni}}{\omega} \Big) F(\omega/\omega_{ti}) -   \frac{\omega_{*Ti}}{\omega} G (\omega/\omega_{ti})  - \frac{N^2(\omega/\omega_{ti})}{D^2(\omega/\omega_{ti})}    \Big]
\end{equation}
Here the characteristic frequencies are $\omega_{*ns} = [(T_s c)/(e_s B)] (\bm{k} \times \bm{b}) \cdot (\bm{\nabla} n)/n $, $\omega_{*Ts} = $ $[(T_s c)/(e_s B)] (\bm{k} \times \bm{b}) \cdot (\bm{\nabla} T_s)/T_s$, $n_i = n$ is the ion density, $\omega_{*ps} = \omega_{*ns}+\omega_{*Ts}$, $\omega_{ti}= \sqrt{2 T_i / m_i}/(q R_0)$, and $F$, $G$, $N$, $D$ are:
\begin{eqnarray}
F(x)  & = & x (x^2 + 3/2) +  (x^4 + x^2 + 1/2) Z(x)\\
G(x)  & = & x (x^4 + x^2 + 2) +  (x^6 + x^4 /2 + x^2 + 3/4) Z(x)\\
N(x)  & = & \Big( 1 - \frac{\omega_{*ni}}{\omega}\Big) [x + (1/2+x^2)Z(x)] \nonumber\\
& & - \frac{\omega_{*Ti}}{\omega} [x(1/2+x^2) + (1/4 + x^4)Z(x)]\\
D(x) & = & \Big( \frac{1}{x}\Big) \Big( 1 + \frac{1}{\tau}\Big) + \Big( 1 - \frac{\omega_{*ni}}{\omega} \Big) Z(x) \nonumber \\
& &  - \frac{\omega_{*Ti}}{\omega}[x + (x^2 - 1/2)Z(x)]
\end{eqnarray}
where $\tau = T_e/T_i$ and $Z(x)$ is the plasma dispersion function:
\begin{equation}
Z(x) = \pi^{-1/2} \int_{-\infty}^{+\infty}\frac{e^{-y^2}}{(y-x)} dy
\end{equation}

The subscript $s$ stands for a particle species index ($s$=i for ions and $s$=e for electrons). The term proportional to $\omega^2 / \omega^2_A$ in $\Lambda$ is the usual polarization current contribution, whereas the other term is due to geodesic curvature coupling.

The GFLDR in the form of Eq.~\ref{eq:GFLDR} does not include FLR and FOW effects, which are proved in Ref.~\cite{Sugama06,Sugama07} to be important. As next steps for this benchmark, we are going to compare results of NEMORB with the extension of the GFLDR given in Ref.~\cite{Zonca98}, and with dispersion relations given in Ref.~\cite{Sugama06,Gao06,Sugama07,Zarzoso12}.

%\end{appendices}

\section*{Acknowledgements}

Enlightening discussions with Bruce Scott, Zhiyong Qiu, Fulvio Zonca and Liu Chen are kindly acknowledged. Fruitful discussions with Guoyong Fu, Garrard Conway and Patrick Simon are also kindly acknowledged.


\begin{thebibliography}{99}
%\vspace{-0.1 cm}
\bibitem{Bottino11} A. Bottino, et al. {\it Plasma Phys. Controlled Fusion} {\bf 53}, 124027 (2011)
\bibitem{Jolliet07} S. Jolliet, et al. {\it Comput. Phys} {\bf 177}, 409 (2007)
\bibitem{Grandgirard08} V. Grandgirard et al. {\it Commun. Nonlinear Sci. Numer. Simul.}, {\bf 13} 81 (2008)
\bibitem{Sarazin10} Y. Sarazin et al. {\it Nucl. Fusion} {\bf 50}, 054004 (2010)
\bibitem{Winsor68} N. Winsor et al. {\it Phys. Fluids} \textbf{11}, 2448, (1968)
\bibitem{Conway11} G. D. Conway et al. {\it Phys. Rev. Letters} {\bf 106}, 065001 (2011)
\bibitem{Fu08} G. Fu et al. {\it Phys. Rev. Lett.} {\bf 101}, 185002 (2008)
\bibitem{Zarzoso12} D. Zarzoso et al. {\it Phys. Plasmas} {\bf 19}, 022102-1 (2012)
\bibitem{Zarzoso13} D. Zarzoso et al.{\it Phys. Rev. Lett.} {\bf 110}, 125002 (2013)
\bibitem{Zonca96} F. Zonca, Liu Chen and R.A. Santoro {Plasma Ph. Control. Fus.} \textbf{38}, 2011-2028 (1996)  
\bibitem{Zonca98} F. Zonca, L. Chen, R.A. Santoro and J.Q. Dong {\it Plasma
Phys. Control. Fus.} {\bf 40}, 2009 (1998)
%\vspace{-0.1 cm}
\bibitem{Sugama06} H. Sugama and T.H. Watanabe, {\it J. Plasma Physics} \textbf{72}, 825 (2006)
%\vspace{-0.1 cm}
\bibitem{Gao06} Zhe Gao, K. Itoh, H. Sanuki, and J. Q. Dong, {\it Phys. of Plasmas} {\bf 13}, 100702 (2006)
\bibitem{Sugama07} H. Sugama and T.H. Watanabe, {\it J. Plasma Physics} \textbf{74}, 139 (2007)
\bibitem{Hallatschek07} K. Hallatscheck, {\it Plasma Phys. Control. Fusion} {\bf 49} B137–B148 (2007)
\bibitem{Angelino08} P. Angelino {\it et al.}, {\it Phys. Plasmas} {\bf 15}, 062306 (2008)
\bibitem{Vernay10} T. Vernay {\it et al.}, {\it Phys. Plasmas} {\bf 17}, 122301 (2010)
\bibitem{Hahm88} T.S. Hahm, W.W. Lee and A. Brizard, {\it Phys. Fluids} {\bf 31}, 1940 (1988)
\bibitem{Brizard07} A.J. Brizard and T.S. Hahm, {\it Rev. of Modern Phys.} {\bf 79} 421 (2007)
\bibitem{Sugama00} H. Sugama, {\it Phys. Plasmas} {\bf 7}, 466 (2000)
\bibitem{Scott10} B. Scott and J. Smirnov, {\it Phys. Plasmas} {\bf 17}, 112302 (2010)
\bibitem{Luetjens96} H. Lütjens, A. Bondeson, and O. Sauter, {\it Comp. Phys. Comm.} {\bf 97}, 219 (1996)
\bibitem{Bottino04} A. Bottino {\it et al.}, {\it Phys. Plasmas} {\bf 11}, 198-206 (2004)
\bibitem{Rosenbluth98} M.N. Rosenbluth and F.L. Hinton, {\it Phys. Rev. Lett.} {\bf 80},4  724 (1998)
\bibitem{Xiao06} Y. Xiao and P. Catto,  {\it Phys. Plasmas} {\bf 13}, 102311 (2006)
\bibitem{Candy04} J. Candy {\it et al.}, {\it Phys. Plasmas} {\bf 11}, L25 (2004)
\bibitem{Zhang10} H.S. Zhang and Z. Lin  {\it Phys. Plasmas} {\bf 17}, 072502 (2010)
\bibitem{Lee83} W. W. Lee, {\it Phys. Fluids} {\bf 26}, 556 (1983)
\bibitem{ScottPC} B. Scott, private communication.
\bibitem{DiTroia12} C. Di Troia, {\it Plasma Phys. Controlled Fusion} {\bf 54}, 105017 (2012)
\bibitem{Qiu10} Z. Qiu, F. Zonca and L. Chen, {\it Plasma Phys. Controlled Fusion} {\bf 52}, 095003 (2010)
\bibitem{Qiu12} Z. Qiu, F. Zonca and L. Chen, {\it Phys. Plasmas} {\bf 19}, 082507 (2012)
\bibitem{Lauber07} Ph. Lauber, S. G\"unter, A. K\"onies, S.D. Pinches, {\it Journal of Comp. Phys.} {\bf 226}, 447 (2007)


\end{thebibliography}
\end{document}